\documentclass[onecolumn, amsmath,amssymb, mathtools, prf]{revtex4-2}

\usepackage{graphicx}
\usepackage{dcolumn}
\usepackage{mathtools}
\usepackage{bm}
\usepackage{cancel}
\usepackage{nicefrac}
\usepackage{multirow}
\usepackage{array}
\usepackage{xcolor}
\usepackage{bigints}
\usepackage{tikz}
\usepackage{cleveref}
\usepackage{bbm}
\usepackage{subcaption}
\usepackage{relsize}
\usepackage{physics, comment}
\usepackage{amssymb}
\usepackage{amsmath}
\usepackage[utf8]{inputenc}
\newcommand{\bsi}{\boldsymbol{\sigma}}
\newcommand{\bol}[1]{\boldsymbol{#1}}

\begin{document}

\title{Enhancement and Suppression of Active Particle Movement Due to Membrane Deformations}

\author{Adam Hitin Bialus}
\email{hitinbialus@mail.tau.ac.il}
\affiliation{School of Physics and Astronomy, Tel Aviv University, Tel Aviv 6997801, Israel.}

\author{Bhargav Rallabandi}
\affiliation{Department of Mechanical Engineering, University of California, Riverside, California 92521, USA}
\altaffiliation{These authors jointly supervised this work.}

\author{Naomi Oppenheimer}
\affiliation{School of Physics and Astronomy and the Center for Physics and Chemistry of Living Systems, Tel Aviv University, Tel Aviv 6997801, Israel.}
\altaffiliation{These authors jointly supervised this work.}

\date{\today}

\begin{abstract}
Microswimmers and active colloids often move in confined systems, including those involving interfaces. Such interfaces, especially at the microscale, may deform in response to the stresses of the flow created by the active particle.
We develop a theoretical framework to analyze the effect of a nearby membrane due to the motion of an active particle whose flow fields are generated by force-free singularities.
We demonstrate our result on a particle represented by a combination of a force dipole and a source dipole, while the membrane resists deformation due to tension and bending rigidity.
We find that the deformation either enhances or suppresses the motion of the active particle, depending on its orientation and the relative strengths between the fundamental singularities that describe its flow. 
Furthermore, the deformation can generate motion in new directions.

\end{abstract}

\maketitle
\footnotetext[1]{These authors jointly supervised this work.}
\section{Introduction}

Interactions between surfaces and viscous flows are abundant both in nature and in engineering~\cite{molaei2014failed,barbish2022dynamics}. A scenario of particular interest occurs when the flow field is produced by an active microswimmer. The coupling between the flow of a microswimmer and fixed boundaries --- either rigid walls or non-deforming fluid interfaces --- has been extensively studied in past years \cite{lauga2006swimming,nasouri2016hydrodynamic,lushi2017scattering,berke2008hydrodynamic,lauga2020fluid}. In such cases, the interactions are limited by the time-reversal symmetry of the Stokes equations \cite{molaei2014failed,frymier1995three,berke2008hydrodynamic}.

More recent work has studied interactions between suspended particles and \emph{deformable} surfaces. The coupling between flow and deformation leads to nonlinear effects, breaking time-reversal symmetry and resulting in rich behavior \cite{trouilloud2008soft,rallabandi2018membrane,montecucco2001living,moriarty2008real,lee2008crawling,dias2013swimming}. Exploration of hydroelastic interactions and motions resulting from them is vital to our understanding of a multitude of different biological and artificial processes, such as the shape of elastic filaments during sedimentation \cite{StoneADupart}, the wake
generated in elastic sheets \cite{arutkin2017elastohydrodynamic,domino2018dispersion,ledesma2016wake},
 the rheology of a suspension of red blood cells \cite{secomb2017blood,guckenberger2018numerical,freund2014numerical,mcwhirter2009flow,pozrikidis2005axisymmetric,secomb1986flow},
the lubrication of joints in limbs \cite{jin2005elastohydrodynamic,dowson1986micro,walker1968boosted}, blood flow in capillaries \cite{dzwinel2003discrete}, and the movement of artificial microswimmers with flexible tails~\cite{dreyfus2005microscopic}. 

When passive particles move near deformable surfaces, fluid-elastic interactions generate additional particle motions, including a ``lift'' of the particle away from the surface \cite{bureau2023lift, rallabandi2024fluid}. These effects have been demonstrated experimentally \cite{rallabandi2018membrane, saintyves2016self, wang2015out} and theoretically  \cite{skotheim2005soft,sekimoto1993mechanism,skotheim2004softl, kargar2021lift, daddi2016long,daddi2017mobility} for a various elastic and viscoelastic surface responses. Careful experimental measurements on passive particles have shown that these fluid-elastic interactions are important down to the nanoscale \cite{davies2018elastohydrodynamic, zhang2020direct, fares2024observation}.  This suggests that similar interactions may be relevant for active motion of microswimmers and synthetic active matter at the microscale. Indeed, a hydroelastic lift was predicted for active swimmers \cite{trouilloud2008soft} near fluid interfaces, while recent work has analyzed other aspects of active swimming near soft interfaces \cite{nambiar2022hydrodynamics, jha2025taylor}. 

In this work, we aim to understand how the presence of a nearby membrane changes the swimming behavior of an active particle. Biological membranes are ubiquitous and their out-of-plane deformations are controlled by a bending rigidity and a membrane tension. The flow generated by a nearby microswimmer can, in principle, produce such deformations, which in turn affect the motion of the swimmer.  We exploit the Lorentz reciprocal theorem to find an analytical expression for the movement of a general class of active particles near deformable membranes. 
We then demonstrate our formalism for several models of active particles. We first consider self-propelled active particles and microorganisms. 
Then we consider active, but not self-propelled, particles (shakers) that can model active proteins near a membrane, such as actin and myosin. In particular, we focus on active particles described by a combination of a symmetric force dipole (a stresslet) and a mass dipole.  
We find that the membrane deformation can either be towards or away from the particle, depending on the orientation of the source and mass dipoles.  
This deformation can either enhance or suppress active motion along both $x$ (parallel to the membrane) and $z$ (perpendicular to the membrane), depending on the relative strength of the source and mass dipoles and their orientation.  Moreover, the deformation can generate new movement both parallel and perpendicular to the membrane.
 We find that for a self-propelled particle, the interactions are inherently long-ranged and depend on the slope of the membrane, whereas for shakers, the interaction depends heavily on the deformation directly below the particle.
For a self-propelled particle, these effects scale linearly with the dipole strength, while for shakers the effect is quadratic in this strength. We study how these interactions vary with the membrane tension relative to its bending rigidity. 

The rest of this paper is organized as follows: in Sec. \ref{section: set up}, we introduce the problem set-up and governing equations. In Sec \ref{sec general integral}, we derive a method of using the reciprocal theorem to extract the velocity correction for small spherical active particles.
Sec. \ref{subsection particle model}, describes the active particle used for our numerical calculations. Sec. \ref{sec self prop} provides results for self-propelled particles, and Sec. \ref{sec shakers} explores shakers in the different limits of the relative strength $q/|D|h$. 
Lastly, in Sec. \ref{section: discussion}, we discuss the results, provide potential applications, and suggest further research directions.

\section{Background}\label{section: Back}
\subsection{Problem Set up}\label{section: set up}
\begin{figure}
    \centering
    \includegraphics[width=0.5\linewidth]{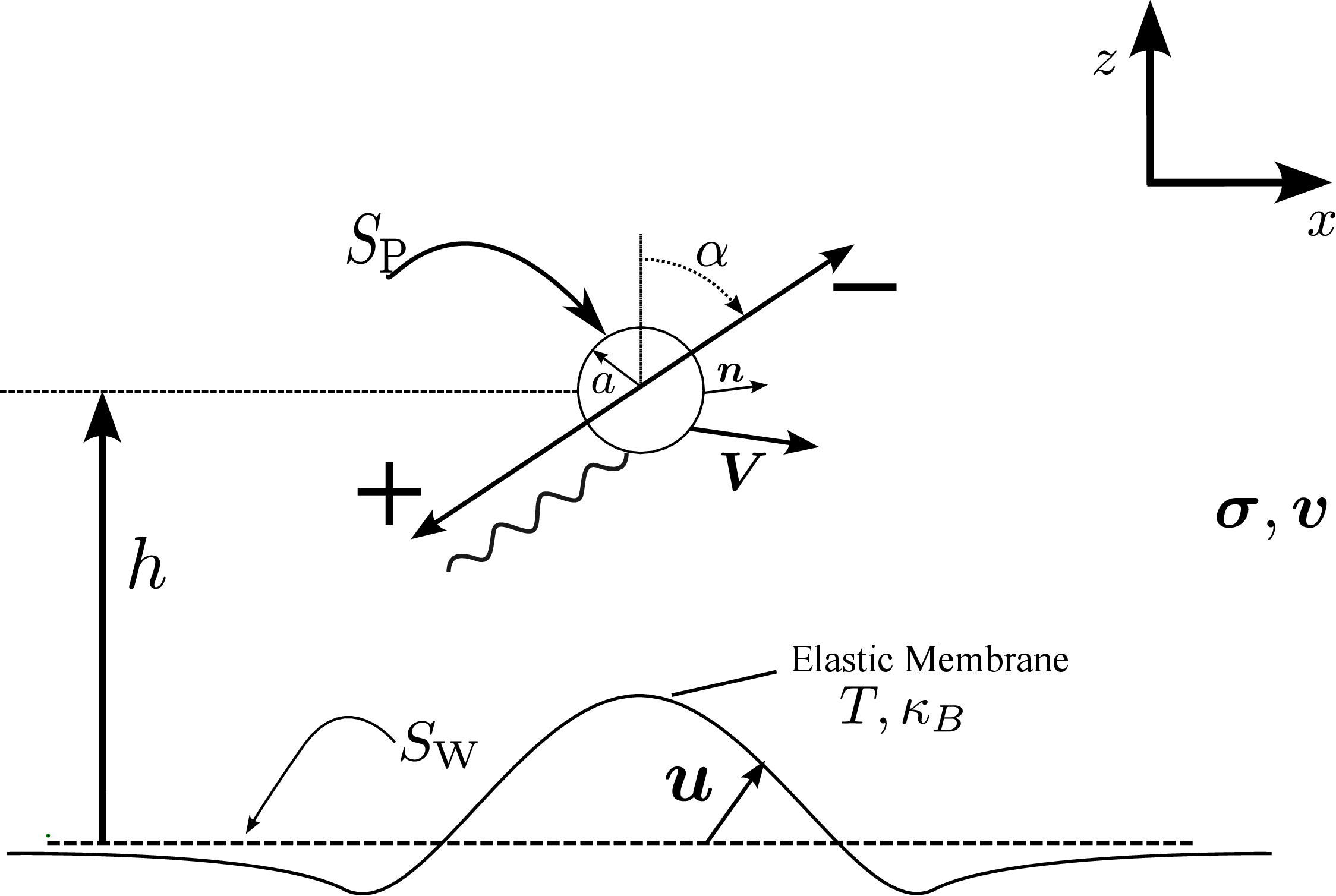}
    \caption{\small A schematic of the problem setup of a microswimmers modeled as a force and mass dipoles oriented at an angle $\alpha$ to an elastic membrane}
    \label{ANGLE}
\end{figure}
We consider a spherical active particle of radius $a$ placed at $\bm{R}=\{0,0,h\}$,  where $h>0$ is the height above an elastic membrane. In its equilibrium state, the membrane is planar and spans the $xy$ plane. The particle translates with velocity $\bm{V}$ in a viscous fluid of viscosity $\eta$. The particle's motion creates a flow that deforms the membrane. The deformation of the membrane then induces a secondary flow field, which in turn affects the motion of the particle. We are interested in solving for this effect under the condition of small membrane deformations. 

In the limit of low Reynolds numbers ($a\rho|\bm{V}|/\eta \ll 1$, with $\rho$ being the fluid density), the velocity field $\bm{v}$ and stress field $\bm{\sigma}$ are governed by the Stokes equations \cite{leal2007advanced,nambiar2022hydrodynamics}
\begin{equation}\label{stokes equations}
    \nabla \cdot \bm{v}  = 0, \quad \nabla \cdot \bsi = \bm{0},
\end{equation}
where $\bsi= -\bm{I}p + \eta \left(\nabla\bm{v} + (\nabla\bm{v})^T\right)$ is the fluid stress tensor produced by the pressure field $p$ and flow field $\bm{v}$ (here $\bm{I}$ is the identity tensor in $\mathbb{R}^3$). 
The flow field due to the active particle will then create deformations in the membrane $\bol{u}(\bm{x}_W, \bm{R}, t)$, where $\bm{x}_W$ represents a point in the plane of the undeformed membrane.  
We focus only on out-of-plane deformation, so $\bm{u} = \{0,0, u_z\}$.
The elastic, no-slip membrane has bending rigidity $\kappa_B$, surface tension $T$, and is initially flat on the $xy$ plane.
Following the linearized Helfrich model, the deformation satisfies \cite{helfrich1973elastic}
\begin{equation}\label{helfrich equation}
    (\kappa_B\nabla^4_{||} - T\nabla^2_{||})u_z = \sigma_{zz}, 
\end{equation}
with $\nabla_{||} = \{\partial_x,\partial_y,0\}$ being the gradient operator in the plane of the undeformed membrane. No-slip conditions on the membrane-fluid interface are
\begin{equation} \label{NoSlip}
    \frac{d}{dt} (\bm{x}_W + \bm{u})\big{|}_{\bm{x}_W} =  \bm{v}\big{|}_{\bm{x}_W + \bm{u}},
\end{equation}
where $\frac{d}{dt}$ represents a material derivative. 

\subsection{General Integral Result}\label{sec general integral}
To find the induced velocity due to the membrane deformation, we invoke a perturbation solution for small membrane deformations and exploit the Lorenz reciprocal theorem. We introduce a dimensionless parameter  $\Lambda$ as the ratio of the characteristic deformation amplitude to the distance of the swimmer from the membrane, i.e. $u_z = O(\Lambda h)$.
We will later define $\Lambda$ in terms of the details of the flow generated by the swimmer, which will generally be dominated by a force dipole.
We then write the flow velocity as $\bm{v} =\bm{v}_0 + \bm{v}_1$, where $\bm{v}_0$ is the flow field of a particle near a flat, rigid, no-slip wall, and $|\bm{v}_1|\ll |\bm{v}_0|$ is the deformation-induced velocity which is a linear function of $\Lambda$.
For the rest of the paper, the subscript 0 refers to quantities of an active particle near a flat, rigid, no-slip wall, whereas the subscript 1 refers to solutions linear in $\Lambda$.
The reciprocal theorem relates the cross-dissipation of energy between two fields, the stress and velocity fields in the problem of interest, and those of a model flow field. 
It provides an integral equation that allows us to extract desired information on the problem at hand, such as force and particle velocity, from a model problem. The model problem we use is that of a Stokes flow produced by a rigid spherical particle of radius $a$ moving with velocity $\hat{\bm{V}}$ either parallel or perpendicular to a flat, no-slip rigid wall located at the $xy$ plane. The induced model velocity and stress fields of the model problem are denoted by $\hat{\bm{v}}$ and $\hat{\bm{\sigma}}$, respectively. 

We expand out the particle swimming velocity $\bm{V}$, the fluid velocity $\bm{v}$ and the stress $\bm{\sigma}$ as 
\begin{align} \label{Expansion}
    (\bm{V}, \bm{v}, \bm{\sigma}) =  (\bm{V}_0, \bm{v}_0, \bm{\sigma}_0) +  (\bm{V}_1, \bm{v}_1, \bm{\sigma}_1) + \dots 
\end{align}
We recall once again that quantities with subscript 1 are linear in the deformation parameter $\Lambda$.
The swimming velocity near a rigid wall is $\bm{V}_0$, which is modified by $\bm{V}_1$ due to the deformation.
Due to linearity, the first-order fields satisfy the Stokes equations \eqref{stokes equations}. 
Since we are interested in the deformation-related contribution to swimming, $\bm{V}_1$, we apply the reciprocal theorem to relate the model flow to the the first-order fields $\bm{v}_1, \bsi_1$, \cite{lorentz1896allgemeiner,kim2013microhydrodynamics,leal2007advanced} 
\begin{equation}\label{recipproblem}
    \int_S \bm{n}\cdot\bm{\sigma}_1  \cdot \hat{\bm{v}} dS  = \int_S \bm{n}\cdot\hat{\bm{\sigma}}  \cdot \bm{v}_1dS.
\end{equation}
Here, the boundary $S$ comprises three distinct parts: the boundary of the particle $S_P$, the boundary at the wall $S_W$, and the boundary at infinity $S_\infty$. The integrals at infinity vanish due to the decay of the flow and stress fields away form the particle. On the left hand side, the integral at the wall, $S_W$, vanishes due to the no-slip condition satisfied by the model flow. Due to the boundary condition on the particle satisfied by the model problem, the contribution of the integral on $S_P$ on the left-hand-side of \eqref{recipproblem} is $\hat{\bm{V}} \cdot \int_{S_P} \bm{n}\cdot\bm{\sigma}_1  dS$, which vanishes identically since the self-propelled particle is force-free. 

We thus focus on the right-hand side of Eq. (\ref{recipproblem}) which yields
\begin{equation}\label{eq olambda}
     -\bm{V}_{1} \cdot\int_{S_P} \bm{n}\cdot\hat{\bm{\sigma}}  dS =\bm{V}_{1} \cdot\hat{\bm{F}} = \int_{S_W} \bm{n}\cdot\hat{\bm{\sigma}}  \cdot \bm{v}_1 dS
\end{equation}
where $\bm{V}_1 = \bm{v}_{1}|_{S_P}$ is the first order correction to the particle's velocity due to the membrane deformation. 
Also, $\hat{\bm{F}}$ is the force exerted by the particle in the model problem on the fluid; the change in sign in the last step stems from the orientation of $\bm{n}$ into the fluid. 


To finish the calculation, we evaluate the right-hand side of Eq. (\ref{eq olambda}).
To that end, we first establish the velocity correction $\bm{v}_1$ at the undeformed wall surface $S_W$ to first order in $\Lambda$.
The no-slip conditions on the membrane, Eq.~\eqref{NoSlip}, provide a link between the flow $\bm{v}$ and the membrane deformation, $\bm{u}$.  We write out the material derivative in terms of partial derivatives and expand out the right-hand side in a Taylor expansion to obtain
\begin{align}
    \bm{v}_0(\bm{x}_W) + \frac{\partial}{\partial t}\bm{u} + \bm{V} \cdot \nabla_{R}\bm{u} = \bm{v}(\bm{x}_W) + \bm{u} \cdot \nabla \bm{v}(\bm{x}_W) + \dots 
\end{align}
where $\frac{d \bm{x}_W}{dt} = \bm{v}_0$ is the velocity at the plane in the underformed configuration, and $(\nabla_R \bm{u})_{ij} = \partial_{R_i}u_j$, $\textbf{R}$ being the position of the particle. We assume that the deformation is advected quasi-statically by the moving particle, and that transient relaxation effects are fast. Invoking the expansion in Eq.~\eqref{Expansion} and retaining all terms up to a linear order in the deformation, we obtain the velocity at the wall as 
\begin{equation}\label{velocity on wall}
    \bm{v}_1|_{\bm{x}_W} = \bm{V}_0 \cdot \nabla_{R}\bm{u} -  \bm{u}\cdot(\nabla \bm{v}_0)|_{\bm{x}_W}.
\end{equation}
The final step is then to solve for the deformation $\bm{u}$ by using Eq. (\ref{helfrich equation}), which to first order in $\Lambda$ is simply
    \begin{equation}\label{eq hell dimensionless}
    (\kappa_B\nabla^4_{||} - T\nabla^2_{||})u_z = \sigma_{0,zz}. 
\end{equation}
Finally, we arrive at the equation for the velocity correction
\begin{equation}\label{general result}
   \bm{V}_{1}\cdot\hat{\bm{F}} = \int_{S_W} \bm{n} \cdot \bm{\hat{\sigma}}  \cdot (\bm{V}_0 \cdot \nabla_{R}\bm{u} - \bm{u}\cdot\nabla \bm{v}_0) dS.
\end{equation}
This equation provides a direct link between the particle's velocity due to the deformation, $\bm{V}_1$, and the other \textit{known} quantities of the problem, namely $\bm{V}_0$, $\bm{v}_0$, $\bm{u}$, and $\bm{\hat{\sigma}}$. 

The above result is general to all force-free active particles, and is therefore applicable to a plethora of active particles and microswimmers, provided that the zeroth order solution near a no-slip wall is known. Furthermore, due to the linearity of Eqs. (\ref{helfrich equation}) and (\ref{stokes equations}), given a swimmer model comprised of a set of $N$ singularities centered at $\bm{R}$, the solution to the velocity correction is readily given by  
\begin{equation}\label{general result N}
   \bm{V}_{1} \cdot\hat{\bm{F}} = \int_{S_W} \bm{n} \cdot \bm{\hat{\sigma}}  \cdot  \sum_{i=1}^N \sum_{j=1}^N \left(\bm{V}_0^i \cdot \nabla_{R}\bm{u}^j - \bm{u}^j\cdot\nabla \bm{v}_0^i\right) dS,
\end{equation} 
where we assume that each singularity translates with velocity $\bm{V}_0^i$, producing velocity and deformation fields  $\bm{v}_0^i$ and $\bm{u}^i$. 
The above equation forms the main result of this paper, and gives a general expression for the velocity of an active force-free particle due to a nearby interfacial deformation. 

\section{Velocity induced by membrane deformation}
In this section we present the solutions to Eq. (\ref{general result N}) for a variety of model particles.
We will first consider an active particle that has a self-propulsion velocity and induces a flow by a force dipole and a source dipole.
As an example, we will take typical values measured for E. coli \cite{drescher2011fluid}.
We will then consider a ``shaker" type of active particle \cite{marchetti2013hydrodynamics}, i.e. a particle that applies active stresses on its environment but is not self-propelled; examples can be active proteins such as myosin. For such cases, we will look at three cases: (a) a force and mass dipoles of equal strengths, (b) when the force dipole is dominant, and (c) when the mass dipole is dominant. 

\subsection{Active Particle Description}\label{subsection particle model}
We first consider an active particle that self-propels with velocity $\bm{V}_{\rm act}$, which is realized in the absence of a membrane.
Near the membrane, the propulsion velocity becomes modified by interactions with the (rigid) wall, as well as corrections from the deformation. The wall does not deform due to $V_{\rm act}$ itself, but by the flow that the particle creates. Since the particle is force-free, its far-away flow field decays as that of a force dipole \cite{drescher2011fluid}.
Furthermore, as the particle moves through the liquid, it displaces the matter in front of it, ``moving it" to the back, resulting in a mass dipole-like behavior \cite{witten2020review}.
Consequently, we represent the particle as a combination of mass and force dipoles \cite{trouilloud2008soft}. 
We solve Eq. (\ref{stokes equations}) in the far-field limit, i.e., $h\gg a$.
In free space, a force dipole generates flow and pressure fields of the form \cite{kim2013microhydrodynamics,leal2007advanced} 
\begin{equation}\label{eq free force}
    v_{i}(\bm{r}) = \frac{D_{jk}}{8\pi\eta}\left(-\frac{r_i\delta_{jk}}{r^3}+3\frac{r_i r_j r_k}{r^5} + \frac{r_k \delta_{ij} - r_j \delta_{ik} }{r^3}\right), \quad p(\bm{r}) = \frac{D_{jk} }{4\pi}\left(\frac{3r_j r_k}{r^5} -\frac{\delta_{jk}}{r^3}\right) ,
\end{equation}
where $\bm{r} = \bm{x}-\bm{R}$ is the relative position vector of a field point $\bm{x}$ and the particle's position $\bm{R}$, and repeated indices are summed over.
Here $\bm{D}$ is a tensor product of an orientation vector $\bm{\nu}_\text{or}$ and a force vector $\bm{\nu}_f$.
For a symmetric force dipole (stresslet) $\bm{D}$ takes the general form $\bm{D} = D( \bol{\nu}_F \bol{\nu}_\text{or} +  \bol{\nu}_\text{or}\bol{\nu}_F)$, where $D$ is a dipole strength and $\bm{\nu}_F$ and $\bm{\nu}_{\text{or}}$ represent force and an orientation vector, respectively. We consider swimmers whose force and orientation vectors are the same, i.e., $\bm{\nu}_\text{or} = \bm{\nu}_F = \{\sin(\alpha),0,\cos(\alpha)\}$, resulting in the dipole tensor
\begin{equation}
\bm{D}=D\begin{pmatrix}
\sin^2(\alpha) & 0 & \sin(\alpha)\cos(\alpha)\\
0 & 0 & 0\\
\sin(\alpha)\cos(\alpha) & 0 & \cos^2(\alpha)
\end{pmatrix},
\end{equation}
where we have chosen the dipole to lie in the $xz$ plane without loss of generality. The angle $\alpha$ represents the orientation of the swimmer with respect to the $z$ axis (Fig. \ref{ANGLE}). 
We name the $D_{11}~(D_{33})$ component of the tensor the parallel (perpendicular) term because for $\alpha = \pi/2\,~(\alpha =0)$ this is the only non-zero term. We call $D_{13}$ and $D_{31}$ the off-diagonal terms.

\begin{figure}
    \centering
    \includegraphics[width=0.5\linewidth]{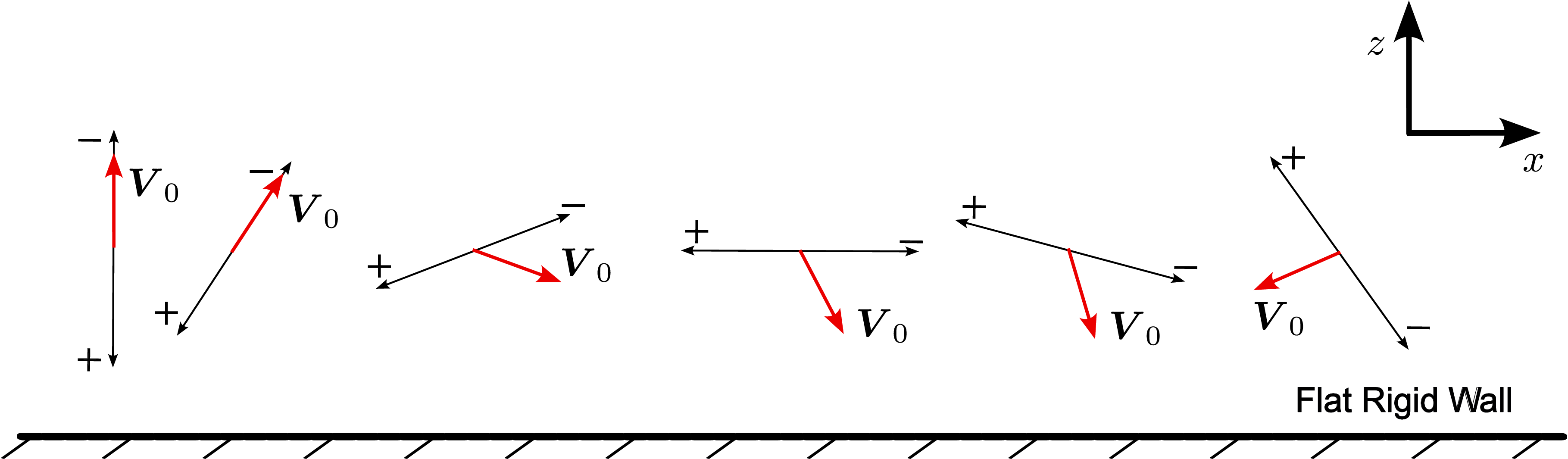}
    \caption{\small A schematic of the induced velocity due to a flat rigid wall $\bm{V}_0$ as a function of incident angle for a model particle}
    \label{fig. flat wall scheme}
\end{figure}

A mass dipole in free space generates a flow
\cite{kim2013microhydrodynamics,leal2007advanced} 
\begin{equation}
     v_{i}(\bm{r}) = \frac{q_{j}}{8\pi\eta}\left(\frac{\delta_{ij}}{r^3}-3\frac{r_i r_j }{r^5}\right), \quad p(\bm{r}) = 0,
\end{equation}
with $\bm{q} = q\sin(\alpha)\bm{e}_x + q\cos(\alpha)\bm{e}_z$.
A flat, rigid wall, alters the flow due to no-slip on the surface of the wall. The flow field solution due to the presence of the flat wall is known for the stresslet and mass dipole and the resultant velocity of the particle is then \cite{blake1974fundamental,gimbutas2015simple} 
\begin{equation}
\begin{aligned}\label{eq total velocity}
    \bm{V}_0 =&  \bm{V}_{\rm act}+\left(\frac{3D_{13}}{32\pi \eta h^2} + \frac{q_1}{32\pi \eta h^3}\right)\bm{e}_x +\left(\frac{3}{64\pi \eta h^2}[2D_{33}-D_{11}] + \frac{q_3}{8\pi \eta h^3}\right)\bm{e}_z \\
    =& \bm{V}_{\rm act}+ \frac{|D|}{\eta \pi h^2}\left[\left(\frac{3\cos(\alpha)\sin(\alpha)}{32} + Q\frac{\sin(\alpha)}{32}\right) \bm{e}_x +  \left(\frac{3}{64}[2\cos^2(\alpha) - \sin^2(\alpha)]+Q\frac{\cos(\alpha)}{8}\right)\bm{e}_z\right],
    \end{aligned}
\end{equation}
where we introduce the dimensionless relative strength of the dipoles $Q =  q/|D|h$. Let us note that the correction to the velocity from the deformation is non-linear in this parameter, and so it is not possible to simply superimpose separate results for the force-dipole and the mass dipole.
As we will show, the combined flows produce a non-trivial correction. Eq. (\ref{eq total velocity}) describes the interaction of the different singularities with a flat, rigid wall.
Even for a pure stresslet or a pure mass dipole, the wall will induce motion. A schematic depiction of the velocity due to a flat wall $\bm{V}_0$ with $\bm{V}_{\rm act} = \bm{0}$ is shown in Fig. \ref{fig. flat wall scheme}.

We wish to understand the effect of the deformation on the particle motion, characterizing both the enhancement or suppression of the motion, as well as the generation of motion in new directions. 
For motion in the $xz$ plane and neglecting in-plane deformations, we evaluate gradients of velocity at the wall and insert it into Eq. (\ref{general result N}) to find 
\begin{equation}\label{eq total correction}
   \bm{V}_{1} \cdot\hat{\bm{F}} = \int_{S_W} \left\{\hat{\sigma}_{zz}  \left(V_{0,z} \pdv{u_z}{h}  -V_{0,x} \pdv{u_z}{x} \right) -  u_z\left(\hat{\sigma}_{zx}\pdv{v_{0,x}}{z}  +\hat{\sigma}_{zy}\pdv{v_{0,y}}{z}\right)\right\}dS.
\end{equation}
We rescale all lengths by a characteristic height $h_0$, and define $d = h/h_0$ to be the dimensionless height. Furthermore, we rescale the velocity and stress fields by a  characteristic velocity $V_P$
\begin{equation*}
    \bm{v} = V_P \bm{v}^*, \quad\text{and}\quad\bm{\sigma} = \frac{V_P \eta}{h_0} \bm{\sigma}^*, \quad
\end{equation*}
where $*$ indicates a dimensionless quantity.
 We then define the dimensionless parameter $\Lambda$, which controls the amplitude of the membrane's deformation, as 
\begin{equation}
    \Lambda = \frac{|D|}{\kappa_B}
\end{equation}
and the dimensionless deformation
\begin{equation}
    u_z = \Lambda h_0 u_z^*,
\end{equation}
with the assumption that $\Lambda \ll1$. 
To evaluate the integral of Eq. (\ref{eq total correction}) numerically, we need to know the stress field of the model problem $\hat{\bsi}$. At the wall, this model flow is dominated by that of a point force (Stokeslet) with $\hat{\bm{F}} = 6\pi\eta a \hat{\bm{V}}$. We thus scale the quantities in the model problem as
 \begin{equation}
     \hat{\bm{v}} = \frac{a\hat{V}}{h_0} \hat{\bm{v}}^*, \quad\text{and}\quad\hat{\bm{\sigma}} = \frac{a\hat{V}_P \eta}{h^2_0} \bm{\sigma}^*,
 \end{equation}
 where $\hat{V} = |\hat{\bm{V}}|$, and the solution  for the above quantities is given by Blake \cite{blake1971note}.
Lastly, we rescale Eq. (\ref{eq total correction}) to write

\begin{equation}\label{eq total solution}
    \bm{V}_{1} \cdot\hat{\bm{V}} = \frac{D V_P}{6\pi \kappa_B  } \int_{S_W} \hat{\sigma}^*_{zz} \left(V_{0,z}^{*} \pdv{u^{*}_z}{d} - V_{0,x}^{*} \pdv{u^{*}_z}{x^*} \right) -  u_z^{*}\left(\hat{\sigma}_{zx}^* \pdv{v^{*}_{0,x}}{z^*}  +\hat{\sigma}_{zy}^* \pdv{v^{*}_{0,y}}{z^*}\right)dS^*.
\end{equation}

To find the deformation, we define the Fourier transform of a function $f(r,\phi)$ on the two-dimensional plane as
\begin{equation}
    \mathcal{F}[f](k,\theta) = \tilde{f}(k,\theta) = \int_0^{\infty}\int_{0}^{2\pi} r f(r,\phi)e^{-irk\cos(\phi-\theta)}drd\phi. 
\end{equation}
A Fourier transform of Eq. (\ref{eq hell dimensionless}) leads to 
\begin{equation}\label{fouier deformation}
    \tilde{u}^*_z = \frac{V_P h_0^2 \eta}{|D|}\frac{\tilde{\sigma}_{0,zz}^* }{k^4 +\tau k^2},
\end{equation}
where we defined the dimensionless tension $\tau=Th^2/\kappa_B$. We evaluate Eq. (\ref{fouier deformation}) exactly and then perform a numerical inverse Fourier transform to find $u_z^*$. We then insert this result into \eqref{eq total solution} and evaluate the integral to obtain the component of  $\bm{V}_{1}$ along a desired direction $\hat{\bm{V}}$; it is convenient to choose $\hat{\bm{V}} = \bm{e}_x$ or $\bm{e}_z$ to obtain velocities along these directions. Finally, we set $d = 1$ without loss of generality (this merely picks the arbitrary length scale $h_0$).  
\begin{figure}[t]
    \centering
    \subfloat[]{\includegraphics[width=0.3\linewidth]{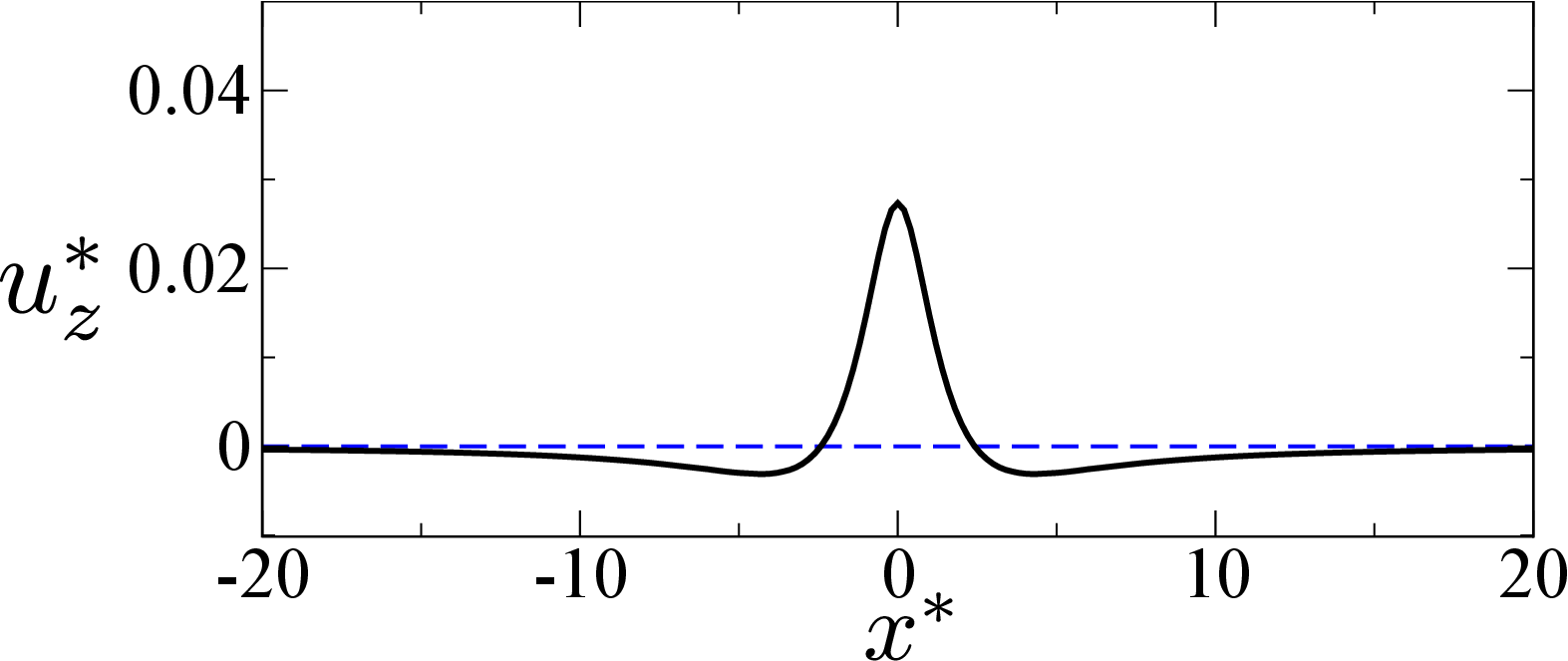}}
    \subfloat[]{\includegraphics[width=0.3\linewidth]{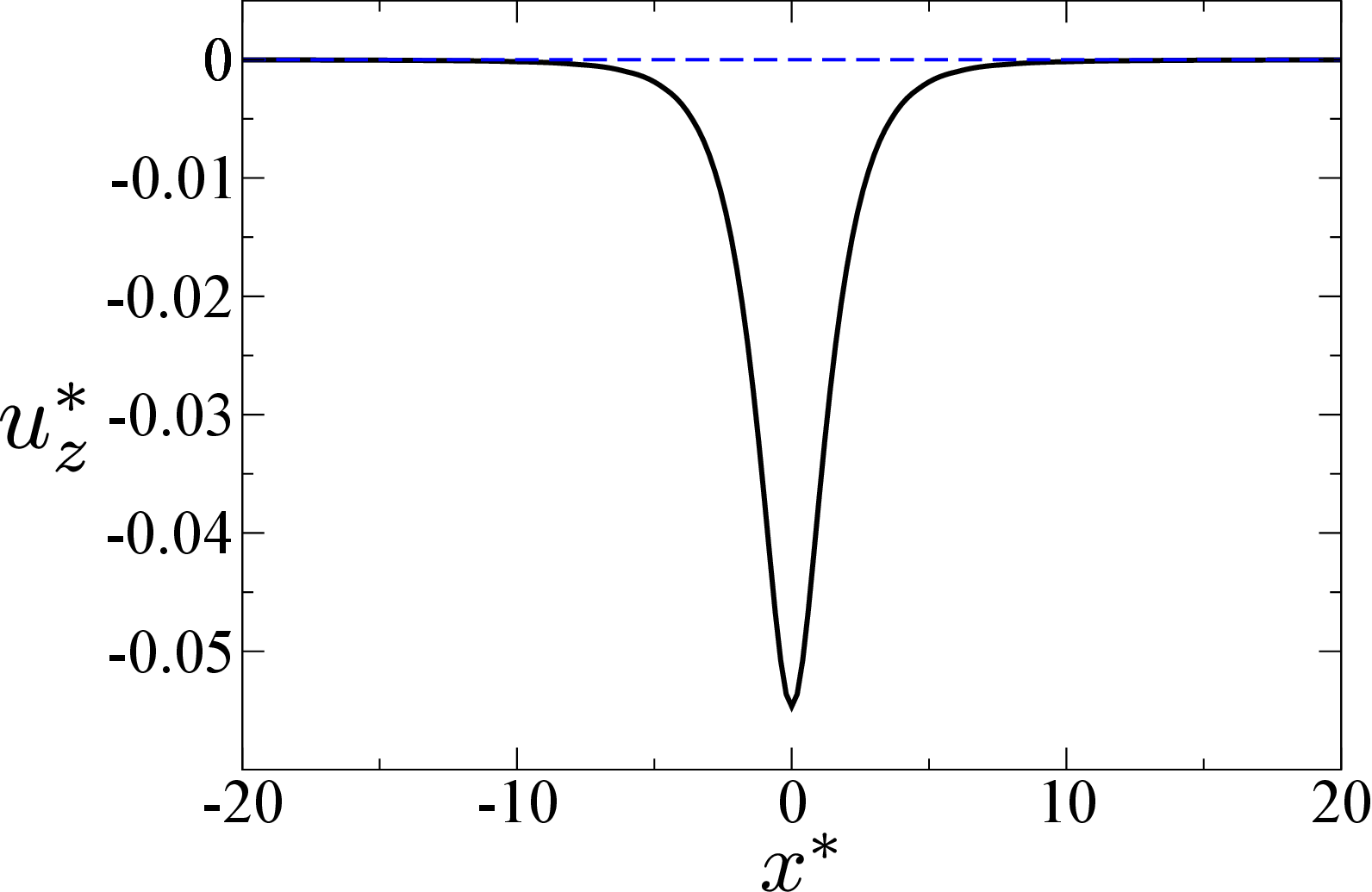}}
    \subfloat[]{\includegraphics[width=0.3\linewidth]{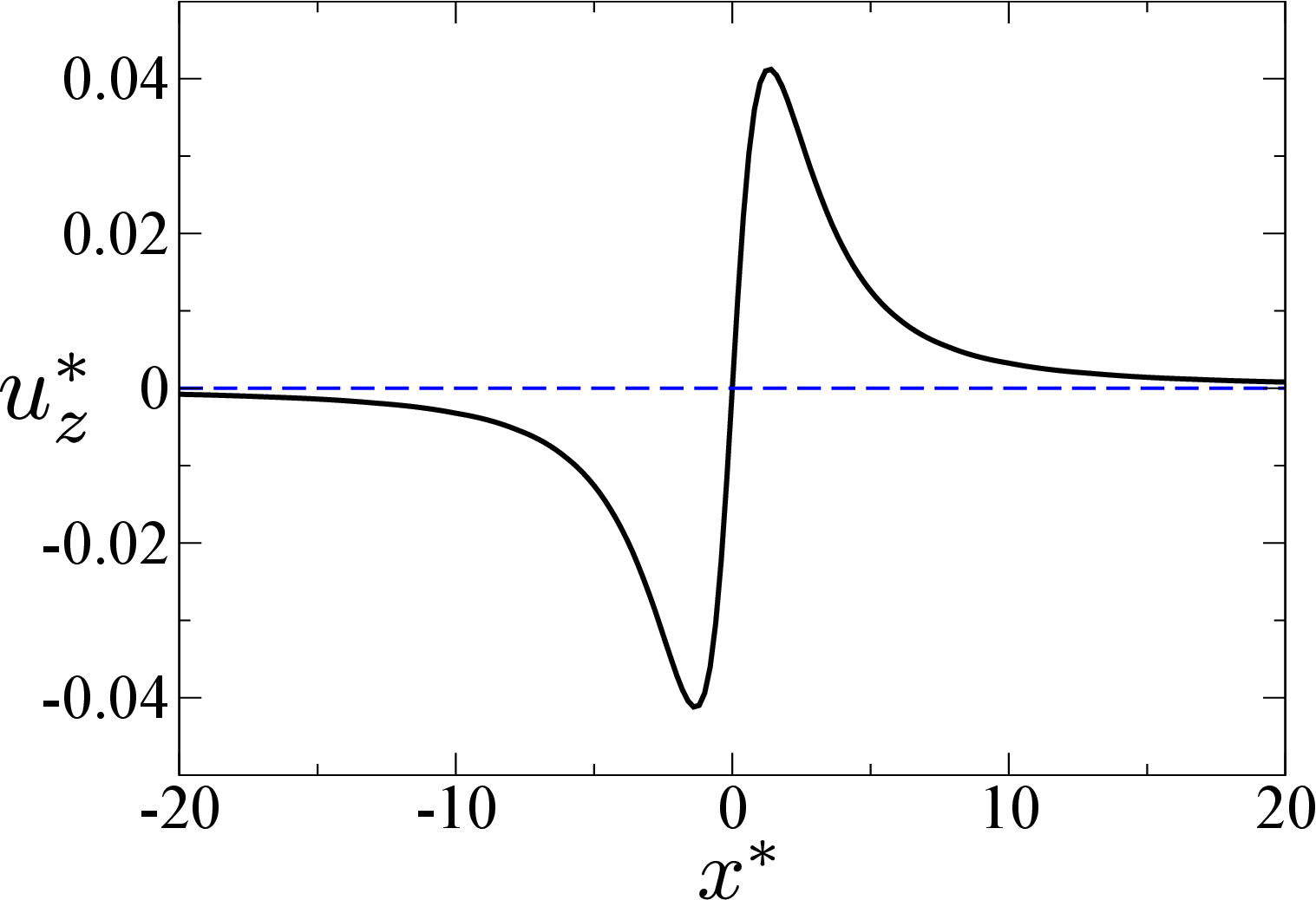}}
    \caption{\small Dimensionless deformation of the membrane due to a stresslet located at $\{0,0,1\}$ with $\tau = 1$ as a function of dimensionless $x^*$ (a) a parallel stresslet (b) a perpendicular stresslet. (c) off diagonal terms. The deformation due to a mass dipole is related by known factors to the deformation due to a stresslet (appendix \ref{app}). \label{fig. force dipole deformations}}
\end{figure}

\subsection{Symmetries}
The deformation due to the stresslet components are presented in Fig. \ref{fig. force dipole deformations}. The deformation due to a mass dipole is related by known factors to the deformation due to a stresslet (appendix \ref{app}). 
Symmetric deformations are produced by the parallel stresslet $D_{11}$, the perpendicular stresslet $D_{33}$, and the perpendicular mass dipole $q_{3}$. Conversely, both the off-diagonal terms, $D_{13}$ and $D_{31}$, and the parallel mass dipole, $q_1$, generate antisymmetric deformations. 
The velocity $\bm{V}_1$ is quadratic in the singularity strengths since it depends both on the stress and the deformation which, in turn, depends on the stress (c.f. Eqs. \eqref{eq total solution} and \eqref{fouier deformation}). We separate these quadratic combinations into self-terms (a result of a single singularity) and cross-terms (interaction between singularities). To provide intuition for which terms contribute to the velocity, we inspect the terms of Eq. (\ref{eq total solution}). The $D_{11},D_{33},\text{ and }q_3$ terms result in a symmetric deformation under $\phi\rightarrow -\phi$ while the deformation due to the $D_{13}$, $D_{31}$, and $q_1$ terms is antisymmetric (table \ref{tab: shcematics}). 
If $\hat{\sigma}_{zz}$ is symmetric in $\phi$, for a given singularity $i$, quadratic self-terms of the form $\hat{\sigma}_{zz} V_z^{i} \partial_h u^{i}_z$ always survive the integration.
Similarly, cross-terms of the form $\hat{\sigma}_{zz} V_z^{i} \partial_h u^{j}_z$ survive the integration only if $i,j$ represent either two symmetric singularities or two antisymmetric singularities.
The same argument holds for terms of the from $u_z^{i}\left(\hat{\sigma}_{zx}\partial_{z} v^{j}_{0,x}  +\hat{\sigma}_{zy}\partial_{z} v^{j}_{0,y}\right)$.
If, on the other hand, $\hat{\sigma}_{zz}$ is antisymmetric, similar arguments show that only cross-terms between symmetric and antisymmetric singularities contribute.

\subsection{Self-propelled particles}\label{sec self prop}
We first consider correction to the velocity of a self-propelled particle ($V_{\rm act} \neq 0$) which can model bacteria such as E. coli, chlamydomonas, or other flagellated swimmers.
We set the characteristic velocity $V_P = V_{\rm act}$. Equation (\ref{eq total solution}) suggests that the velocity correction scales as $V_1 \sim \frac{V_{\rm act} D}{\kappa _B}$. Interestingly, this scale is independent of the distance $h$, suggesting long-range interactions. We will see below that the dependence on $h$ is weak (logarithmic) when the tension is small.  We rely on typical systems and set the ratios \cite{drescher2011fluid}
\begin{equation}
    \frac{D}{\eta a^2} \approx 25 |\bm{V}_{\rm act}|, \text{ and } \frac{h_0}{a} \approx5.
\end{equation}
We let $Q =1$, such that the three terms contributing to the velocity of Eq. \eqref{eq total velocity} are of comparable magnitudes.

\begin{figure}[h]
    \subfloat[]{\includegraphics[width=0.4\linewidth]{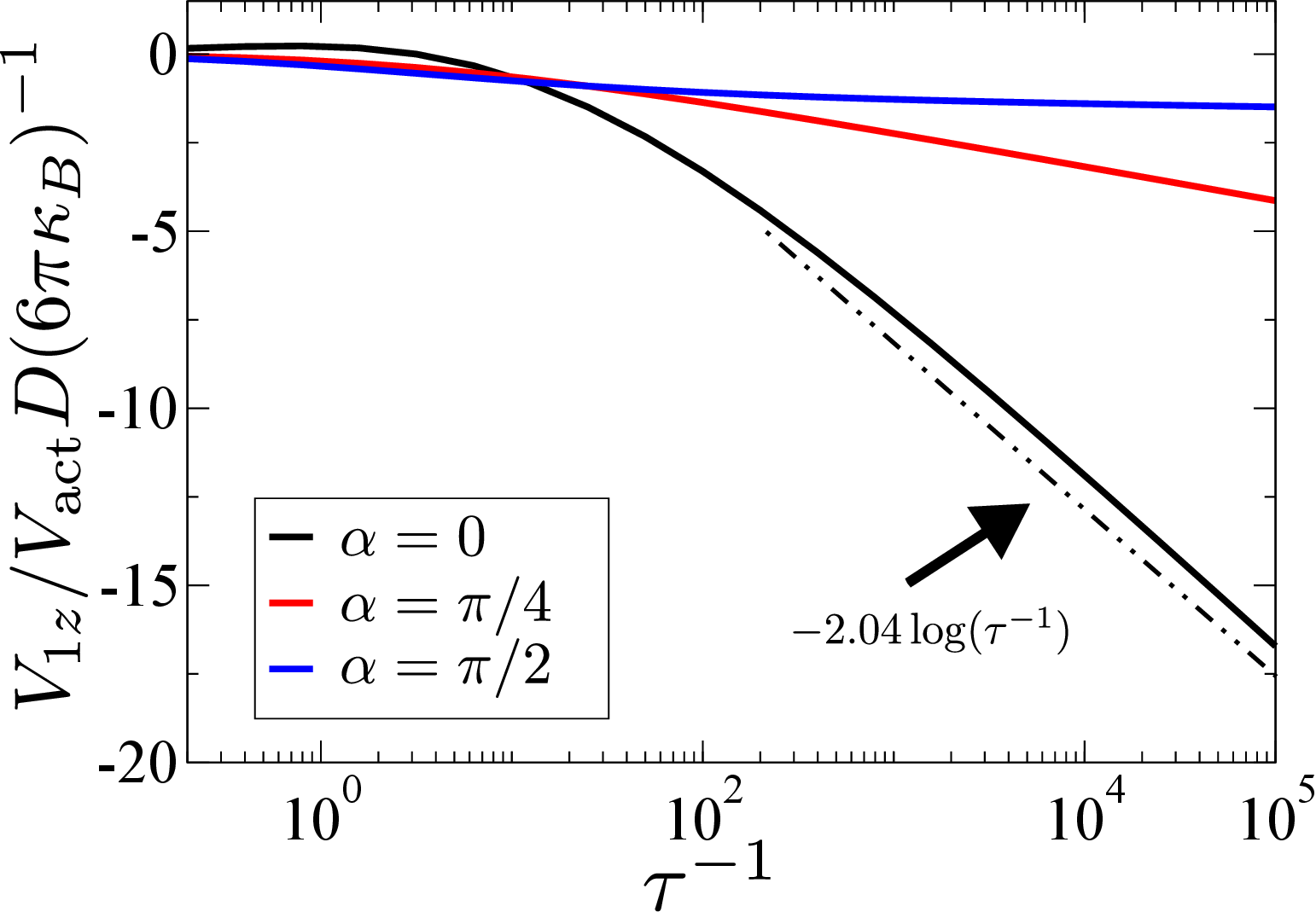}}
    \subfloat[]{\includegraphics[width=0.4\linewidth]{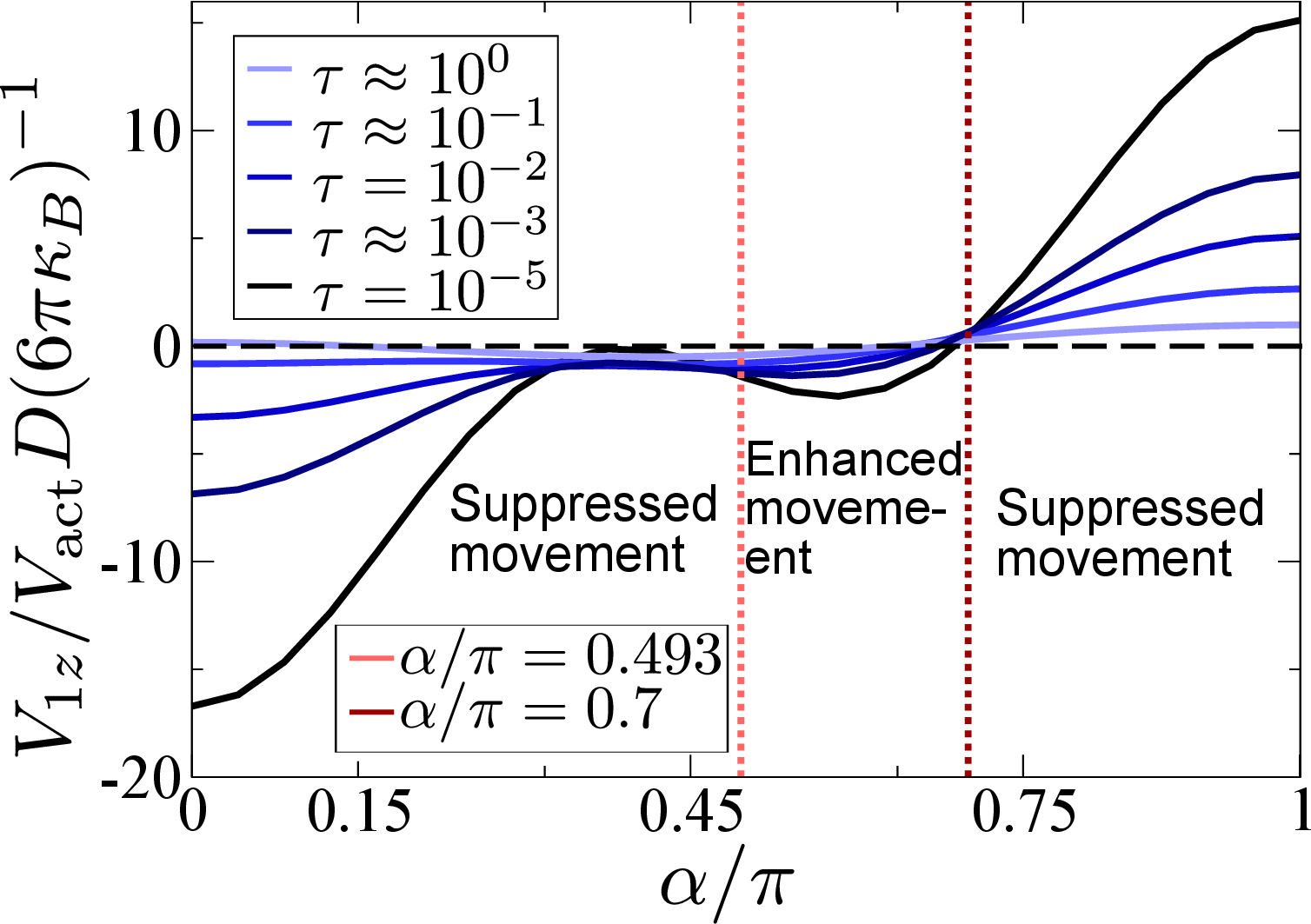}}
    \caption{\small (a) Rescaled induced velocity along $z$ of a self-propelled particle with force and mass dipole combination ($Q =1$) as a function of dimensionless tension $\tau$. Blue line is swimming parallel to the membrane, black line is perpendicular to membrane, and red is swimming in $\alpha = \pi/4$. (b) Rescaled induced velocity along $z$ of a self-propelled particle with force and mass dipole combination ($Q =1$) as a function of orientation angle $\alpha$.\label{fig. total V in z2}}
\end{figure}

\textbf{Correction to the normal velocity.}
The induced velocity $V_1$ along the $z$ direction (perpendicular to the membrane) as a function of dimensionless tension $\tau$ is presented in Fig. \ref{fig. total V in z2}(a). 
For small membrane tension, the correction scales as $V_1 \sim \log(\tau^{-1})$ (dashed line).
Since $\tau = Th^2/\kappa_B$, we have $V_1 \sim \log(\sqrt{\kappa_B/T}/h)$, where $\sqrt{\kappa_B/T}$ is a length scale set by a competition between bending and tension. 
Asymptotic expansions in this limit are tabulated in table \ref{tab: excat results z}. For larger tensions, the velocity becomes smaller, and scales by an additional power of $h^{-2}$. 
The vertical velocity has a complicated dependence on the orientation angle $\alpha$; see  Fig. \ref{fig. total V in z2}(b). From \eqref{eq total velocity} we find that the particle moves away from the membrane for $\alpha < 0.493 \pi$ and toward it for $\alpha > 0.493 \pi$. This motion is suppressed by deformation both for small angles ($\alpha < 0.493 \pi$) and for large angles ($\alpha > 0.7 \pi$). At intermediate angles ($0.493 \pi < \alpha < 0.7 \pi$) the deformation enhances the motion towards the membrane.

\begin{figure}[h]
    \subfloat[]{\includegraphics[width=0.4\linewidth]{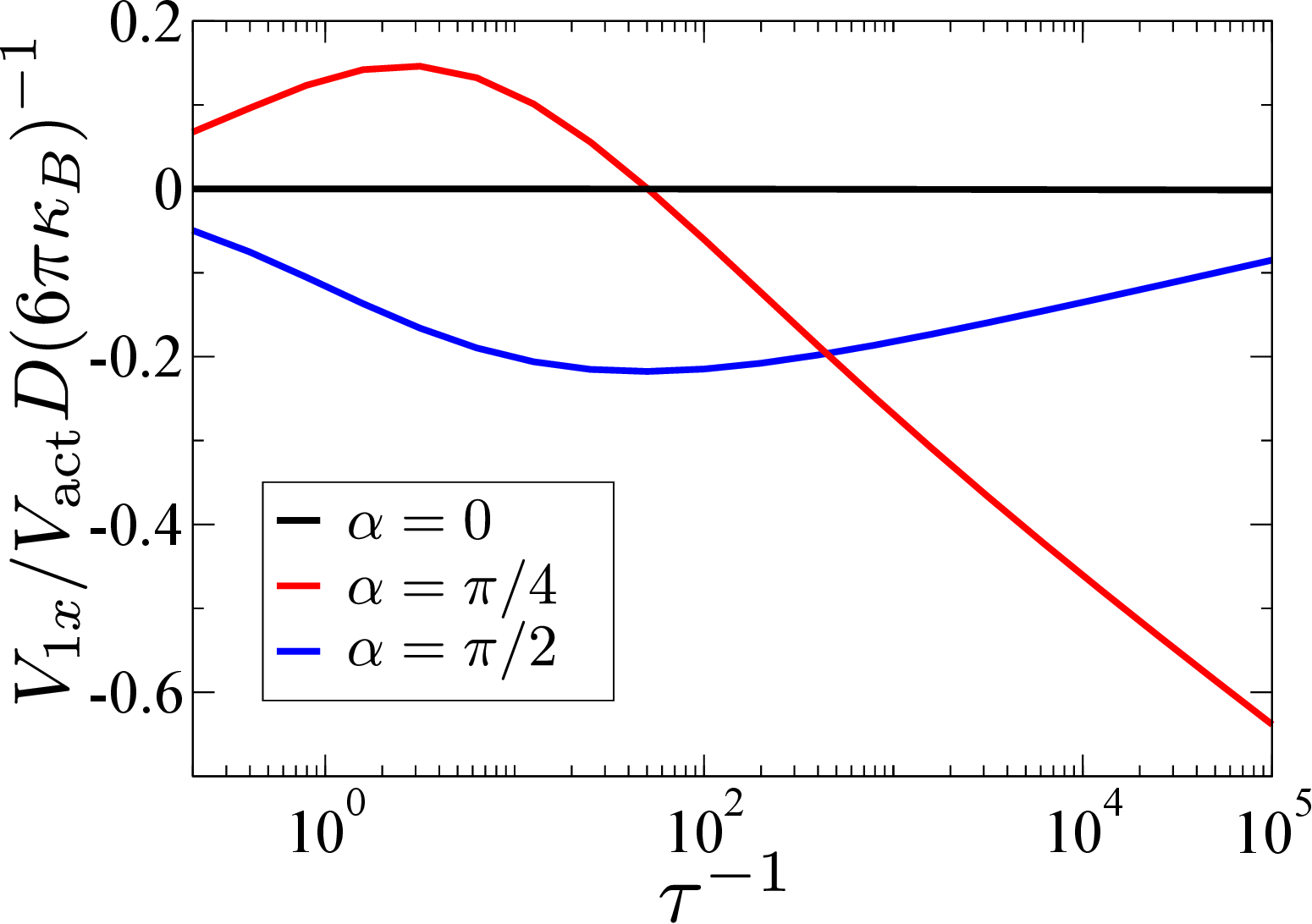}}
    \subfloat[]{\includegraphics[width=0.4\linewidth]{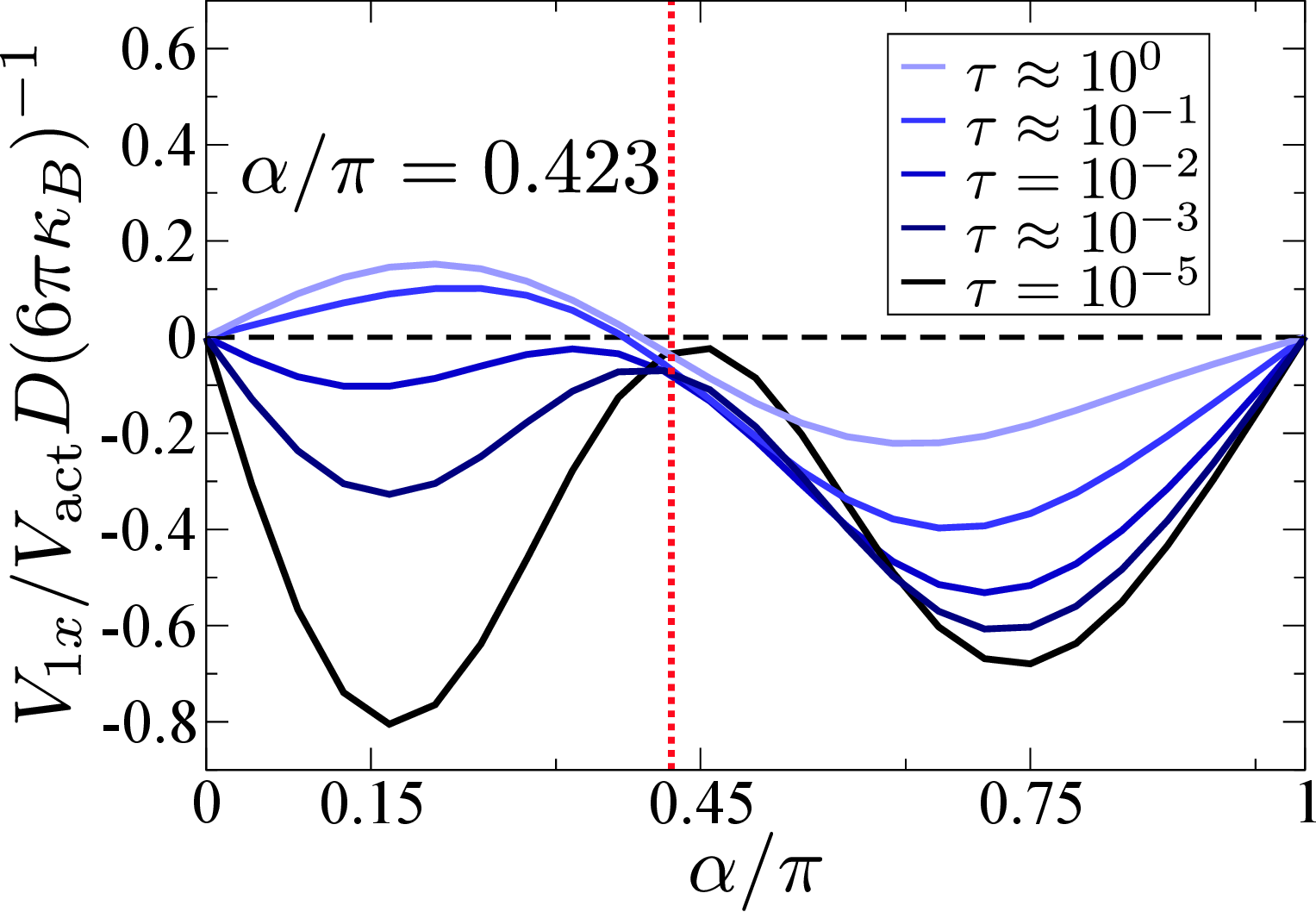}}
    \caption{\small (a) Rescaled induced velocity along $x$ of a self-propelled particle with force and mass dipole combination ($Q =1$) as a function of dimensionless tension $\tau$. Blue line is swimming parallel to the membrane, black line is perpendicular to membrane, and red is swimming in $\alpha = \pi/4$. (b) Rescaled induced velocity along $x$ of a self-propelled particle with force and mass dipole combination ($Q =1$) as a function of orientation angle $\alpha$}.\label{fig. total V in x2}
\end{figure}

\textbf{Correction to the tangential velocity.} 
The particle may either slow down or speed up along $x$ due to the deformation, depending on $\alpha$ and $\tau$ (Fig. \ref{fig. total V in x2}). As before, the correction scales as $\log \tau^{-1}$ for small tensions (see table \ref{tab: excat results x} for more details). For $0 < \alpha < \pi$, the particle moves along $+x$ in the absence of deformation. For small membrane tensions, the deformation retards the self-propulsion along $x$ for all orientations $\alpha$. Conversely, for large tensions ($\tau \gtrsim 10^{-1}$), the deformation enhances motion when $0 < \alpha < 0.42 \pi$, while it suppresses motion along $x$ for larger orientation angles $0.42 \pi <  \alpha < \pi$. 

\subsection{Shakers}\label{sec shakers}
A shaker is an active particle that may drive flow but does not self-propel.
Examples include active proteins near membranes.
However, the flows that they create can interact hydrodynamically with nearby boundaries --- including deformable ones --- leading to ``induced'' motion of a shaker \cite{mikhailov2015hydrodynamic}. Here we demonstrate the effects of the membrane on an active, but not self-propelled particles ($\bm{V}_{\rm act} = \bm{0}$). We find that when the active velocity is negligible, the dominant contribution to the correction comes from the deformation directly below the particle.
If the membrane is deformed towards the particle, its velocity will be enhanced, while if the deformation is away from the particle, it will be suppressed.
In addition to the orientation angle and the dimensionless tension, the motion is controlled by the ratio of the mass and force dipole strengths, $Q$.
We discuss these effects below, focusing on $Q=1$ with further analysis provided in the Appendix.

The natural velocity scale, which we also choose to be the characteristic velocity for non-dimensionalization, is set by the dipole strength as $V_P = |D|/\eta h_0^2$. 
From Eq.~\eqref{eq total solution}, the velocity due to the deformation scales as $V_1 \sim \frac{ D^2}{\eta \kappa_B h_0^2}$, up to a dimensionless function of $\tau$. 

\begin{figure}[h]
    \subfloat[]{\includegraphics[width=0.4\linewidth]{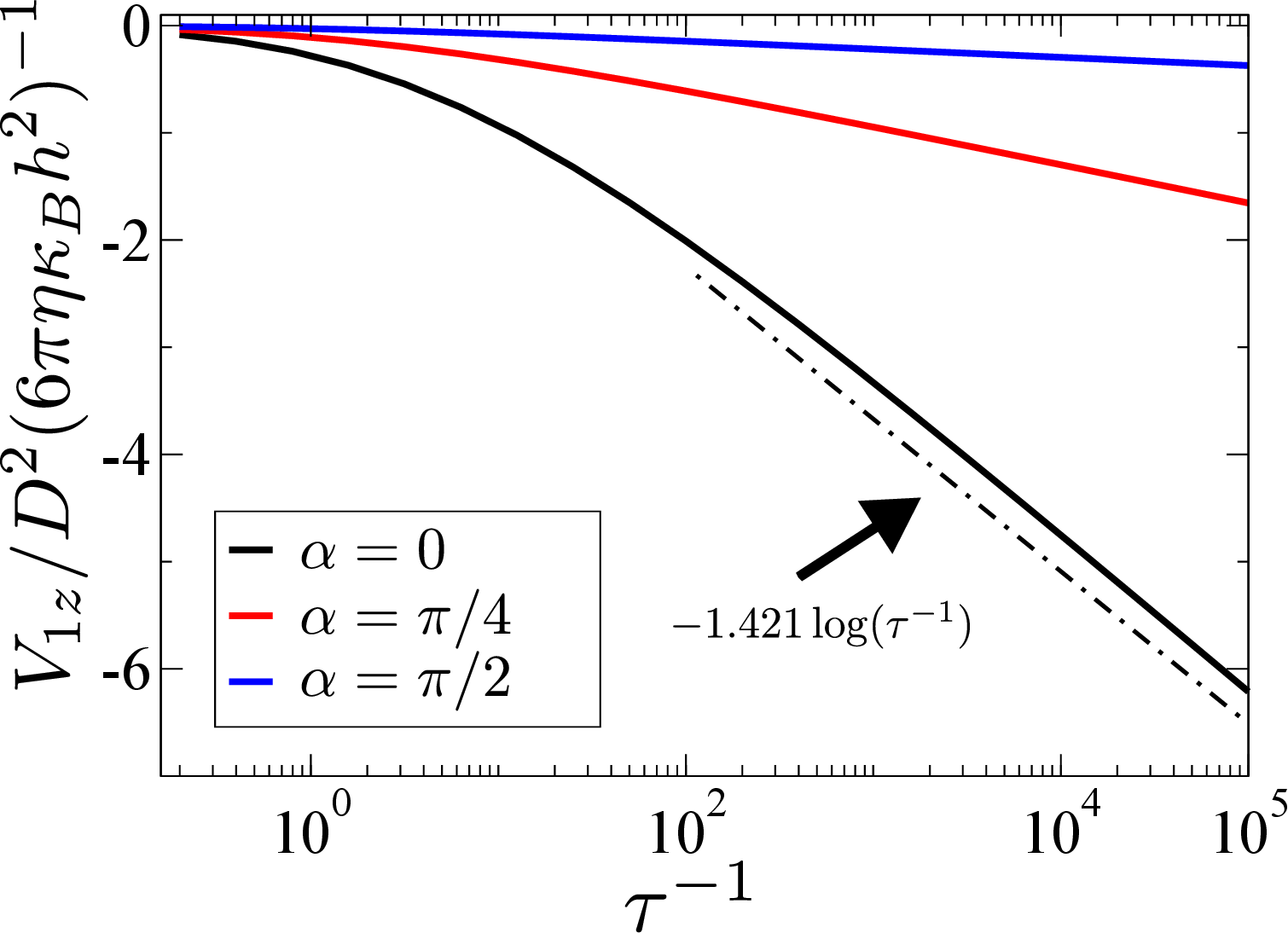}}
    \subfloat[]{\includegraphics[width=0.4\linewidth]{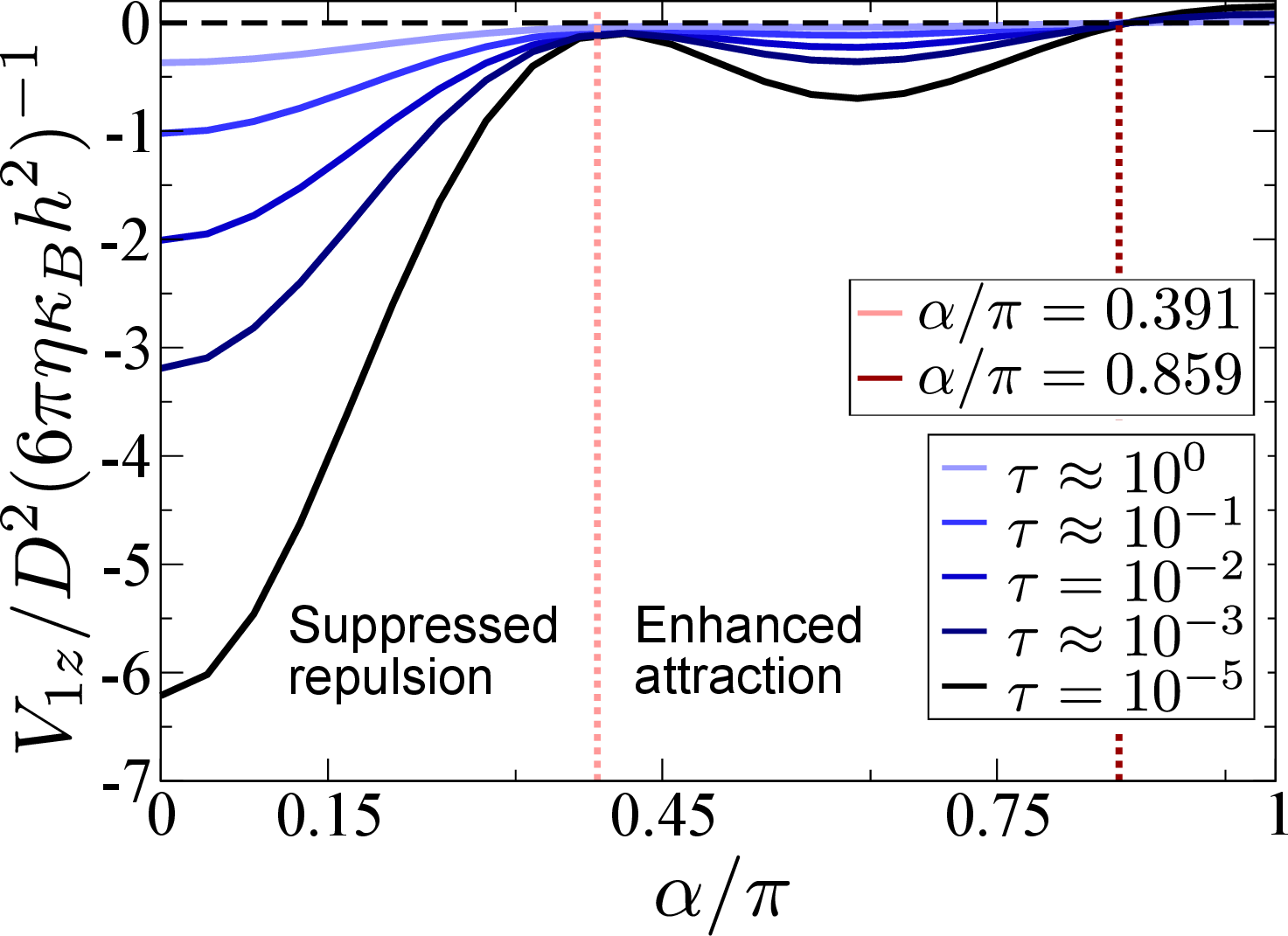}}
    \caption{\small (a) Rescaled induced velocity along $z$ of a force and mass dipole combination ($Q =1$) as a function of dimensionless tension $\tau$. Blue line is swimming parallel to the membrane, black line is perpendicular to membrane, and red is swimming in $\alpha = \pi/4$. (b) Rescaled induced velocity along $z$ of a force and mass dipole combination ($Q =1$) as a function of orientation angle $\alpha$.\label{fig. total in z2}}
\end{figure}

\textbf{Correction to the normal velocity.} 
The induced velocity $V_1$ along the $z$ direction as a function of dimensionless tension $\tau$ is presented in Fig. \ref{fig. total in z2}(a). For small surface tension, the velocity scales as $V_1 \sim \log\tau/h^2$ (See table \ref{tab: excat results z} for details). For most angles and tensions, the membrane deformation leads to a velocity correction towards the membrane. The exception occurs at $\alpha$ close to $\pi$, where the induced velocity is away from the membrane. 
We understand these behaviors by analyzing some limiting cases. For small angles ($\alpha \approx 0$), the particle is repelled from a rigid wall due to hydrodynamic interactions, according to  Eq. (\ref{eq total velocity}). Meanwhile, the flow due to the particle is directed away from the particle towards the membrane (it is roughly that of a perpendicular stresslet), causing the membrane to deform away from the particle (similar to the situation depicted in Fig. \ref{fig. force dipole deformations}b). We can therefore think of the situation with a deformable membrane as ``moving the flat wall further away," which from Eq. (\ref{eq total velocity}) suggests a suppressed repulsion for small $\alpha$ (see Fig. \ref{fig. total in z2}b).  The reverse occurs when $\alpha = \pi/2$. The particle is attracted to a rigid wall according to Eq.~\eqref{eq total velocity}. Furthermore, the flow is dominated by a parallel stresslet and so the membrane deformation is towards the particle, as indicated in Fig.~\ref{fig. force dipole deformations}a. Thus, we can think of the wall as ``moving closer" to the particle as a consequence of deformation, strengthening the interaction and enhancing the attraction towards the membrane.

With these observations, we expect that the crossover between enhancement and suppression occurs when the membrane deformation right below the particle, i.e. at the origin, vanishes. We denote that deformation due to the perpendicular stresslet and mass dipole by $(u^{\perp}_z)$, and the deformation due to the parallel stresslet as $(u^{||}_z)$; the parallel mass dipole does not deform the membrane at the origin. 
Thus, the deformation at the origin is 
\begin{equation}\label{eq def at origin}
     u_z|_{x=y=0} = \left(\left[\cos^2(\alpha) + Q\cos(\alpha)\right]u^{\perp}_z +\sin^2(\alpha)u^{||}_z\right)\bigg |_{x=y=0}.
\end{equation}
From Eqs. (\ref{eq: force pa def}), (\ref{eq: force pe def}), and (\ref{eq source pe deforamtion}), we find that $u_z^{\perp} = -2u_z^{||}$ at the origin. Setting the deformation to zero at the origin provides us with an approximation for $\alpha_{\rm cross}$ 
\begin{equation}\label{eq cross total}
     \cos(\alpha_{\rm cross})= \frac{-Q + \sqrt{Q^2 + 3}}{3}.
\end{equation}

\begin{figure}
    \centering
    \subfloat[]{\includegraphics[width=0.4\linewidth]{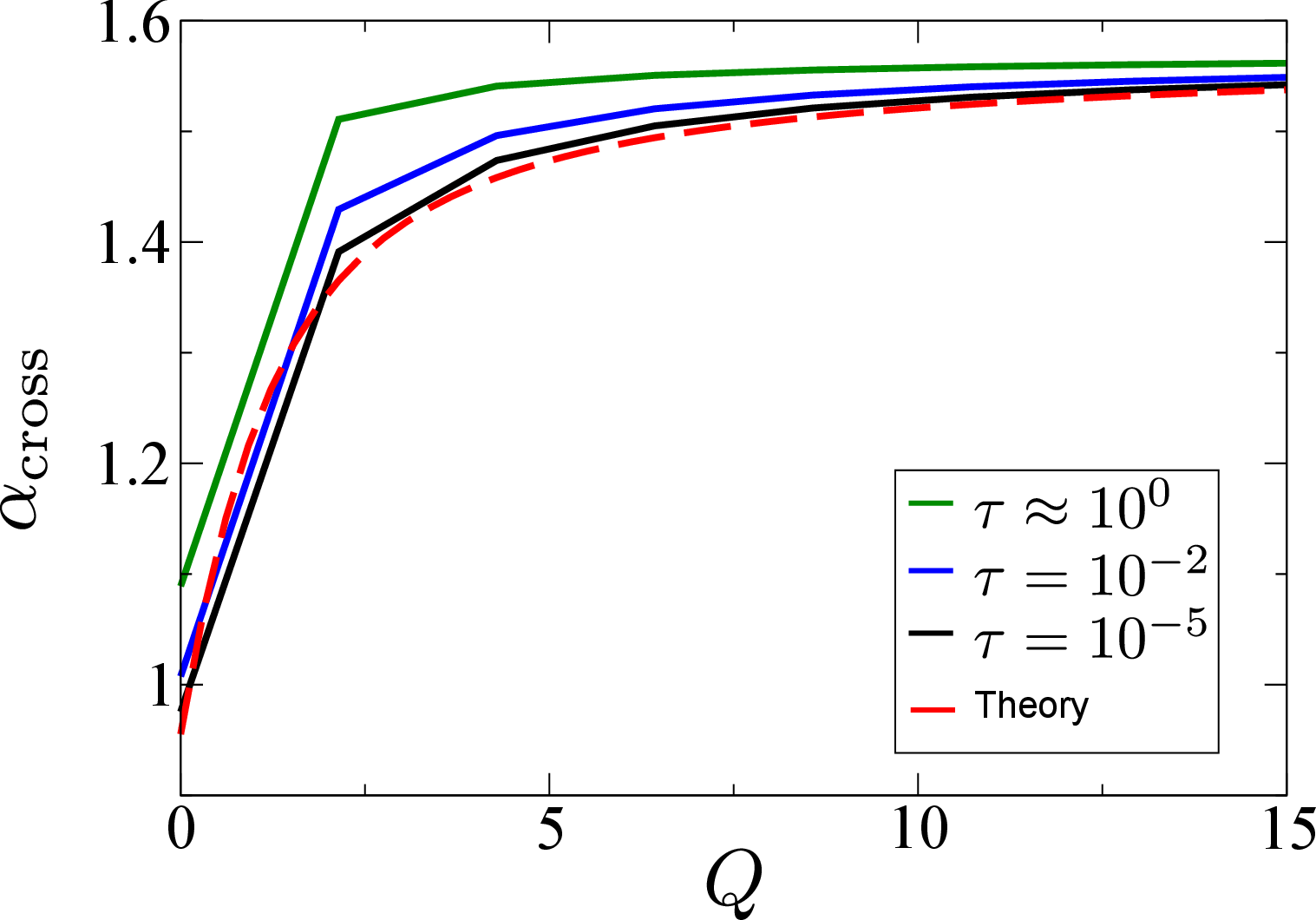}}
    \caption{\small A plot of the crossing angle $\alpha_{\rm cross}$ for a force and mass dipole combination as a function the relative strength $Q$\label{theory comp}}
\end{figure}

For  $Q =1$,  we obtain a prediction $\alpha_{\rm cross}= 1.23 \approx 0.4 \pi$ from Eq.~\eqref{eq cross total}. This is consistent with the detailed calculation in Fig. \ref{fig. total in z2}(b), which shows the velocity correction along $z$ as a function of the incident angle $\alpha$; the predicted $\alpha_{\rm cross}$ is indicated by the leftmost vertical red line.
Indeed, at $\alpha_{\rm cross}$, the correction is small, and it stems from the antisymmetric part of the deformation alone, which is zero at the origin, suggesting that the deformation at the origin is the most significant contributor to the velocity correction. Fig. \ref{theory comp} shows $\alpha_{\rm cross}$ as a function of the ratio $Q$. The crossover angle is relatively insensitive to the tension $\tau$, which is consistent  with the prediction of Eq.~\eqref{eq cross total}, which depends only on the ratio $Q = q/|D| h$. 

Lastly, for $\alpha/\pi \approx 0.859$ (rightmost vertical line in Fig. \ref{fig. total in z2}(b)), the deformation-induced velocity changes sign and becomes positive. We attribute this behavior to a competition between the perpendicular terms of the stresslet and mass dipole, which either ``push" or ``pull" the particle away from or toward the membrane, but scale differently with $h$. 



%

\textbf{Correction to the tangential velocity.} The tangential velocity due to the deformation can be either positive or negative depending on the angle. As before, we understand the results from the symmetries of the different terms, and from the shape of the deformation. For moderate $\alpha$, it follows a similar pattern of suppressed or enhanced movement as the normal velocity, with a transition occurring for $\alpha/\pi \approx 0.391$. Interestingly, at $\alpha/\pi\approx0.607$, deformation induces particle motion, even though the particle does not move along $x$ at this angle in the rigid case.

\begin{figure}
    \centering
    \subfloat[]{\includegraphics[width=0.4\linewidth]{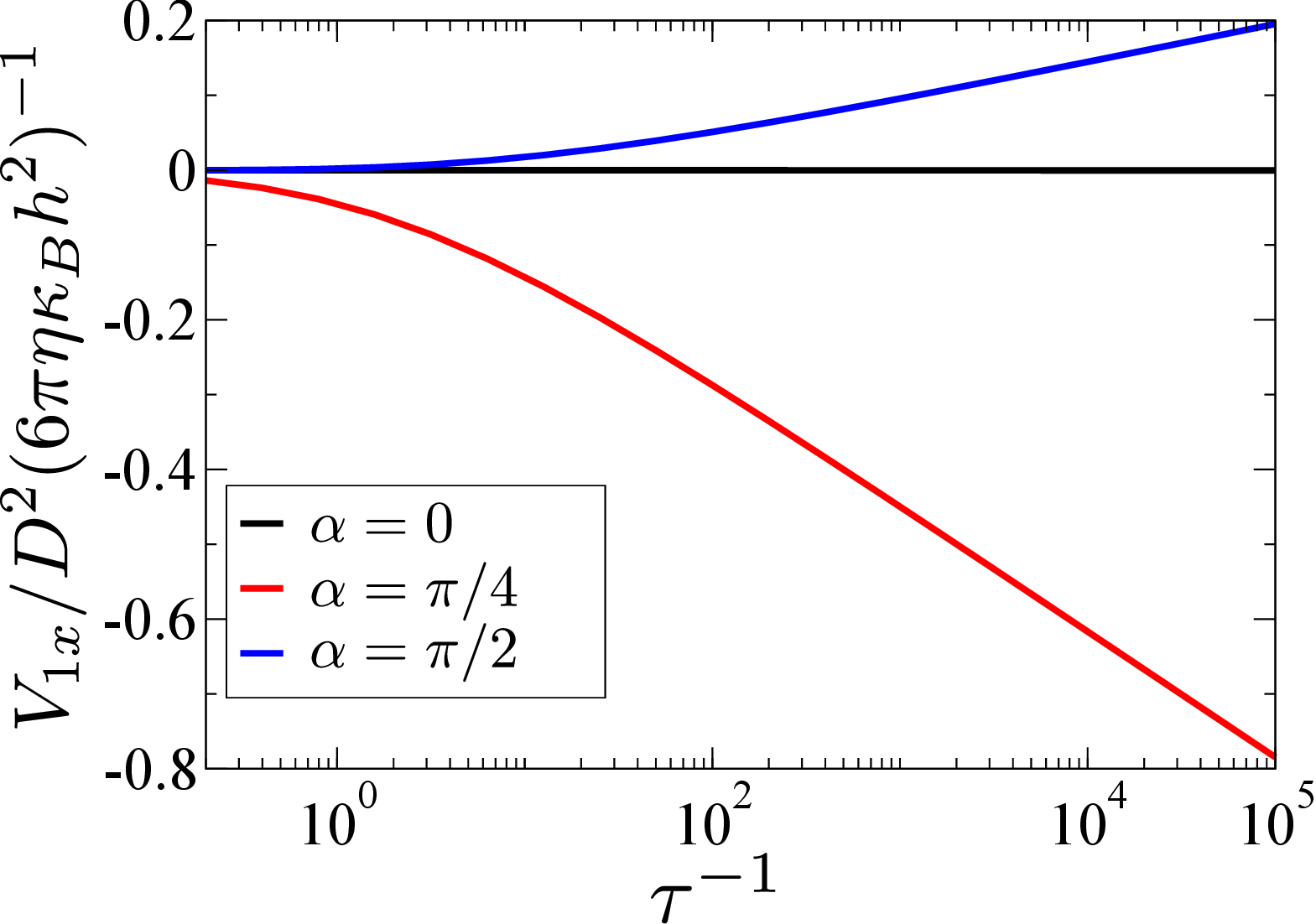}}
    \subfloat[]{\includegraphics[width=0.4\linewidth]{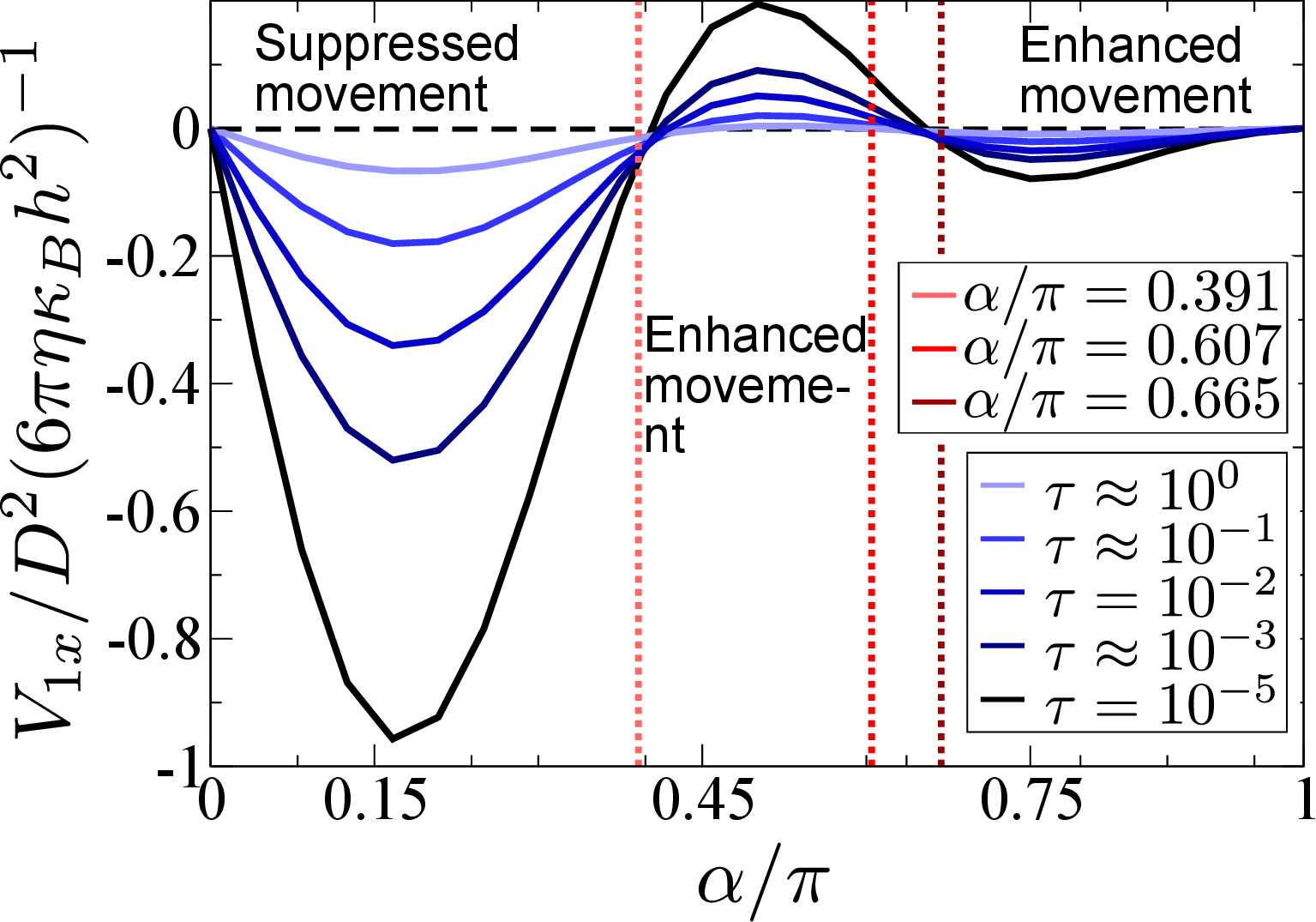}}
    \caption{\small (a) Rescaled induced velocity along $x$ of a force and mass dipole combination ($Q =1$) as a function of dimensionless tension $\tau$. Blue line is swimming parallel to the membrane, black line is perpendicular to membrane, and red is swimming in $\alpha = \pi/4$. (b) Rescaled induced velocity along $x$ of a force and mass dipole combination ($Q =1$) as a function of orientation angle $\alpha$.\label{fig. total in x}}
\end{figure}

\textbf{Other limits of dipole and mass strengths.} The regions of enhanced and suppressed motion depend on the dipole strength ratio $Q$.
For small $Q$, the force dipole is dominant and the crossover angle becomes approximately $0.304 \pi$ (c.f Fig. \ref{theory comp} with $Q=0$).
The deformation induces a correction towards the surface for all $\alpha$.
When $\alpha < \alpha_{\rm cross}$, the deformation suppresses the interaction relative to the undeformed case (repulsion), while for $\alpha > \alpha_{\rm cross}$, it enhances the interaction (attraction).
Motion along $x$ mirrors the same pattern of enhancement and suppression, with a positive $V_{1x}$ for $\alpha < \alpha_{\rm cross}$ and negative $V_{1x}$ for $\alpha > \alpha_{\rm cross}$. 
When $Q\rightarrow\infty$, the force dipole is negligible and the mass dipole dominates.
It is now necessary to choose the characteristic velocity as $V_P = q(\eta h_0^3)^{-1}$ leading to $\Lambda = q(h_0\kappa_B)^{-1}$.
Therefore, the correction velocity due to the deformation now scales as $V_1 \sim \frac{q^2}{\eta \kappa_B h^4}$.
In this case, the crossover angle according to \eqref{eq cross total} is $\alpha_{\rm cross} = \pi/2$ (c.f Fig. \ref{theory comp} with $Q=15$).
Along both normal and tangential directions, the effect of deformation from the mass dipole mirrors that of the stresslet, but with the modified crossover angle. 
For a more detailed analysis of the effect of $Q$ on the correction, see appendices \ref{subsection: force} and \ref{subsection: source}.

\section{Discussion}\label{section: discussion}
In this work we used the reciprocal theorem to derive an analytical expression for the first-order correction to the velocity of an active particle in the presence of an infinite elastic membrane with bending rigidity and surface tension. 
Our approach relies on far-field effects and small membrane deformation while allowing for arbitrary orientation with respect to the membrane, as long as no net forces act on the particle. 
We first present a particle modeled as a force and mass dipole combination with relative strength $Q =1$ and active velocity $V_{\rm act}$. 
We show the deformation can either enhance or suppress the motion of the particle associated with interactions with a rigid wall, depending on the incident angle $\alpha$. 
When $V_{\rm act}$ is negligible we develop an analytic expression to predict the value where the effect changes from suppression to enhancement and show that it agrees well with the detailed calculation. 
Furthermore, we find that deformation can generate movement along both $z$ and $x$ directions for certain incident angles. 

The approach developed here can be used on a plethora of model swimmers provided no net force is applied and the solution near a rigid no-slip wall is known.
These calculations can be useful to better understand the behavior in a variety of biological microswimmers such as E. coli, or even robotic swimmers.
The effects of non-uniform bending rigidity of the membrane, e.g. due to accumulation of cholesterol in the veins, can also be considered \cite{chakraborty2020cholesterol}.


\section*{Acknowledgments}
This work was supported by NSF-BSF Grant No. 2023624 and NSF Grant No. 2328628.
\bibliography{main}

\appendix*
\section{Calculation of velocity components and deformations}\label{app}
\subsection{Force dipole}\label{app. force dipole}
We wish to numerically solve the integral in Eq. (\ref{eq total solution}).
We first consider a pure force dipole. The rescaled velocity derivatives are~\cite{gimbutas2015simple}
\begin{equation}
\begin{aligned}
    \partial_{z^*} v^{*}_{0,x}|_{s_w} =& \begin{cases}
        -\frac{3 r \cos (\phi )}{2 \pi } \left(\frac{2}{\left(r^2+1\right)^{5/2}}-\frac{5 r^2 \cos ^2(\phi )}{\left(r^2+1\right)^{7/2}}\right)\sin^2(\alpha), \quad \text{parallel swimming}\\
        -\frac{3 r \cos (\phi )}{2 \pi } \left(\frac{2}{\left(r^2+1\right)^{5/2}}-\frac{5 }{\left(r^2+1\right)^{7/2}}\right)\cos^2(\alpha), \quad \text{perpendicular swimming}\\
        \frac{3 }{2 \pi } \left(\frac{r^2 \cos^2 (\phi)+1}{\left(r^2+1\right)^{5/2}}-\frac{10r^2\cos^2 (\phi ) }{\left(r^2+1\right)^{7/2}}\right)\cos(\alpha)\sin(\alpha), \quad \text{off diagonal terms}\\
    \end{cases}
\\
\\
\partial_{z^*} v^{*}_{0,y}|_{s_w} =&
\begin{cases}
     -\frac{3 r \sin(\phi )}{2 \pi } \left(\frac{2}{\left(r^2+1\right)^{5/2}}-\frac{5 r^2 \cos ^2(\phi)}{\left(r^2+1\right)^{7/2}}\right)\sin^2(\alpha), \quad \text{parallel swimming}\\
     -\frac{3 r \sin (\phi )}{2 \pi } \left(\frac{2}{\left(r^2+1\right)^{5/2}}-\frac{5 }{\left(r^2+1\right)^{7/2}}\right)\cos^2(\alpha), \quad \text{perpendicular swimming}\\
     \frac{3\cos(\phi)\sin(\phi)r^2 }{2 \pi } \left(\frac{1}{\left(r^2+1\right)^{5/2}}-\frac{10}{\left(r^2+1\right)^{7/2}}\right)\cos(\alpha)\sin(\alpha), \quad \text{off diagonal terms},
\end{cases}
\end{aligned}
\end{equation}
where we defined $x = r\cos(\phi),y=r\sin(\phi)$ as the coordinates on the flat wall and we took $d=1$.
Next, to calculate the correction along $z$, the model problem is a stokeslet pointing along $z$ where the stress field $\hat{\bsi}^*$ is ~\cite{blake1971note,blake1974fundamental}
\begin{equation}
    \hat{\sigma}^*_{zz} = \frac{9}{\left(r^2+1\right)^{5/2}}.
\end{equation}
Due to the symmetry of the model problem we can write $\hat{\sigma}^*_{zx} = \hat{\sigma}^*_{zr}\cos(\phi)$ and $\hat{\sigma}^*_{zy} = \hat{\sigma}^*_{zr}\sin(\phi)$. Moreover, since $w_r = w_x \cos(\phi) + w_y \sin(\phi)$ for any vector $\bol{w}$, the second terms in Eq. (\ref{eq total solution}) can be written as
\begin{equation}
    u_z^{*}\hat{\sigma}_{zr}^*\partial_{z^*} v^{*}_{0,r}, 
\end{equation}
with
\begin{equation}
    \hat{\sigma}_{zr}^* = -\frac{9 r}{\left(r^2+1\right)^{5/2}}.
\end{equation}
Lastly, we write the deformation in Fourier space
\begin{equation}\label{eq: force pa def}
    \tilde{u}_z^*(k,\theta) = \frac{d k^2 e^{-d k} \cos ^2(\theta )}{k^4+\tau k^2  }\sin^2(\alpha) \quad \text{parallel swimming},
\end{equation}

\begin{equation}\label{eq: force pe def}
    u_z^*(k,\theta) = -\frac{d k^2 e^{-dk }}{k^4+k^2 \tau }\cos^2(\alpha) \quad \text{perpendicular swimming},
\end{equation}
and 

\begin{equation}
    u_z^*(k,\theta) = -\frac{2 i d k^2 e^{-k q} \cos (\theta )}{k^4+k^2 \tau }\cos(\alpha)\sin(\alpha) \quad \text{Off diagonal terms},
\end{equation}
and use Mathematica to numerically perform the inverse transform.
Notice that at $r = 0$, the inverse Fourier transform of Eq. (\ref{eq: force pa def}) and (\ref{eq: force pe def}) are identical up to a factor of $-2$ which stems from the negative sign and the integration over $\cos^2(\theta)$. This implies that at the origin the amplitude of the deformation due to the perpendicular term will be twice that of the parallel term.

We can understand which terms will survive the integration by a closer inspection of Eq. (\ref{eq total solution}). 
We separate the interaction into quadratic terms (a result of a single singularity) and cross-terms (interaction between singularities). 
Notice that the parallel and perpendicular terms result in a symmetric deformation under $\phi\rightarrow -\phi$ while the deformation due to the off-diagonal terms is antisymmetric (cf. table \ref{tab: shcematics}). 
Moreover, $\partial_h u_z$ never breaks this symmetry while $\partial_x u_z$ reverses it. 
Now, since $\hat{\sigma}_{zz}$ is symmetric in $\phi$, for a given singularity $i$, quadratic terms of the form $\hat{\sigma}_{zz} V_z^{i} \partial_h u^{i}_z$ always survive the integration.
Similarly, cross-terms of the form $\hat{\sigma}_{zz} V_z^{i} \partial_h u^{j}_z$ only survive the integration if $i,j$ represent either two symmetric singularities, or two antisymmetric singularities.
The same argument holds for terms of the from $u_z^{i}\left(\hat{\sigma}_{zx}\partial_{z} v^{j}_{0,x}  +\hat{\sigma}_{zy}\partial_{z} v^{j}_{0,y}\right)$.

The calculation of velocity along the $x$ direction is similar. Here we choose the model problem to be a stokeslet pointing along positive $x$. Now the stress tensor $\hat{\bsi}$ is \cite{blake1971note,blake1974fundamental}
\begin{equation}
    \hat{\sigma}_{zz} = -\frac{9 r \cos (\phi )}{\left(r^2+1\right)^{5/2}}, \quad \hat{\sigma}_{zx} = \frac{9 r^2 \cos^2 (\phi )}{\left(r^2+1\right)^{5/2}}, \text{ and }\hat{\sigma}_{zy} = \frac{9 r^2 \cos (\phi )\sin(\phi)}{\left(r^2+1\right)^{5/2}},
\end{equation}
where now we no longer have symmetry under $\phi \rightarrow \phi + \pi$ and we need both $\hat{\sigma}_{zx}$ and $\hat{\sigma}_{zy}$.
Notice that now the model problem is antisymmetric under $\phi\rightarrow -\phi$. 
A symmetry argument now shows that all quadratic terms will cancel and only cross-terms between symmetric and antisymmetric singularities will survive the integration of Eq. (\ref{eq total solution}).

\subsection{Mass Dipole}\label{app: mass}
For a mass dipole, the rescaled velocity derivatives are \cite{blake1974fundamental}
\begin{equation}
\begin{aligned}
\partial_{z^*} v^{*}_{0,x}|_{s_w} =&
\begin{cases}
    -\frac{3}{2 \pi } \left(\frac{5 r^2 \cos ^2(\phi )}{\left(r^2+1\right)^{7/2}}-\frac{1}{\left(r^2+1\right)^{5/2}}\right), \quad \text{parallel swimming}\\
    \frac{3 r \cos (\phi )}{2 \pi }\left(\frac{5}{\left(r^2+1\right)^{7/2}}-\frac{1}{\left(r^2+1\right)^{5/2}}\right), \quad \text{perpendicular swimming}
\end{cases}
\\
\\
    \partial_{z^*} v^{*}_{0,y}|_{s_w} =& 
    \begin{cases}
        -\frac{15 r^2 \sin (\phi ) \cos (\phi )}{2 \pi  \left(r^2+1\right)^{7/2}}, \quad \text{parallel swimming}\\
    = \frac{3 r \sin (\phi )}{2 \pi }\left(\frac{5}{\left(r^2+1\right)^{7/2}}-\frac{1}{\left(r^2+1\right)^{5/2}}\right) , \quad \text{perpendicular swimming}\\
    \end{cases}
\end{aligned}
\end{equation}
The deformation terms are
\begin{equation}\label{eq source pe deforamtion}
    u_z^*(k,\theta) = -\frac{k^2 e^{-d k}}{k^4+k^2 \tau }\cos(\alpha) \quad \text{perpendicular swimming}.
\end{equation}
Note that this is similar to the perpendicular terms of the force dipole, but with one less factor of $d$,
and 
\begin{equation}
    u_z^*(k,\theta) =-\frac{i k^2 e^{dk} \cos (\theta )}{k^4+k^2 \tau }\sin(\alpha) \quad \text{parallel swimming}.
\end{equation}
A similar symmetry argument as given in Sec. \ref{app. force dipole} shows which terms will survive the integration of Eq. (\ref{eq total solution}).
The same argument extends to cross-terms between the force and mass dipoles.

  \begin{table}[]
        \centering
        \begin{tabular}{|c|c|c|c|}
                \hline
         Direction&Singularity  & symmetry under $\phi\rightarrow\phi +\pi$& correction direction and 
         scaling\\
         \hline
         \multirow{17}{4em}{$z$}&Parallel force dipole & + & $V_{1,z} \sim \frac{D^2}{\kappa_B\eta h^2}(0.148\log(h\sqrt{T/\kappa_B})  +0.04)$ 
         \\
         &Perpendicular force dipole & + & $V_{1,z} \sim \frac{D^2}{\kappa_B\eta h^2}(0.594\log(h\sqrt{T/\kappa_B})  +0.19)$ 
         \\
         &Off diagonal terms & - & $V_{1,z} \sim \text{slower than } \frac{D^2}{\kappa_B\eta h^2}\log(h\sqrt{T/\kappa_B})$ 
         \\
         &Force at $\alpha=\pi/4$ & - & $V_{1,z} \sim \frac{D^2}{\kappa_B\eta h^2}(0.04\log(h\sqrt{T/\kappa_B})  -0.032)$ 
         \\
         &Parallel mass dipole& - & $V_{1,z} \sim \text{ slower than} \frac{q^2}{\kappa_B\eta h^3}\log(h\sqrt{T/\kappa_B})$ 
         \\
         &Perpendicular mass dipole& + & $V_{1,z} \sim \frac{q^2}{\kappa_B\eta h^4}(0.792\log(h\sqrt{T/\kappa_B})  +0.25)$ 
         \\
         &Mass dipole at $\alpha = \pi/4$& - & $V_{1,z} \sim \frac{q^2}{\kappa_B\eta h^4}(0.398\log(h\sqrt{T/\kappa_B})  +0.11)$ 
         \\
          &Parallel force-mass dipoles& - & $V_{1,z} \sim \frac{D^2}{\kappa_B\eta h^2}(0.152\log(h\sqrt{T/\kappa_B})  +0.008)$ 
         \\
         &Perpendicular force-mass dipoles& + & $V_{1,z} \sim \frac{D^2}{\kappa_B\eta h^2}(2.842\log(h\sqrt{T/\kappa_B})  +0.91)$ 
         \\
         &Force-mass dipoles at $\alpha = \pi/4$& - & $V_{1,z} \sim \frac{D^2}{\kappa_B\eta h^2}(0.702\log(h\sqrt{T/\kappa_B})  +0.105)$ 
         \\
         &Parallel self propulsion& + & $V_{1,z} \sim \text{ slower than} \frac{V_{\rm act}D}{\kappa_B }(\log(h\sqrt{T/\kappa_B}))$ 
         \\
         &Perpendicular self propulsion& + & $V_{1,z} \sim \frac{V_{\rm act}D}{\kappa_B }(4.048\log(h\sqrt{T/\kappa_B})  +6.859)$ 
         \\
         &Self propulsion at $\alpha = \pi/4$& - & $V_{1,z} \sim \frac{V_{\rm act}D}{\kappa_B }(0.811\log(h\sqrt{T/\kappa_B})  +0.608)$  
         \\
         \hline
        \end{tabular}
        \caption{Asymptotic results for  velocity correction along $z$ of different singularities}
        \label{tab: excat results z}
    \end{table}

      \begin{table}[]
        \centering
        \begin{tabular}{|c|c|c|c|}
                \hline
         Direction&Singularity  & symmetry under $\phi\rightarrow\phi +\pi$& correction direction and 
         scaling\\
         \hline
         \multirow{17}{4em}{$x$}&Parallel force dipole & + & $V_{1,x} = 0$  
         \\
         &Perpendicular force dipole & + & $V_{1,x} = 0$ 
         \\
         &Off diagonal terms & - & $V_{1,x} = 0$  
         \\
         &Force at $\alpha=\pi/4$ & - & $V_{1,x} \sim \frac{D^2}{\kappa_B\eta h^2}(0.05\log(h\sqrt{T/\kappa_B})  +0.004)$ 
         \\
         &Parallel mass dipole& - & $V_{1,x} = 0$ 
         \\
         &Perpendicular mass dipole& + & $V_{1,x} = 0$  
         \\
         &Mass dipole at $\alpha = \pi/4$& - & $V_{1,x} \sim \frac{q^2}{\kappa_B\eta h^4}(0.1\log(h\sqrt{T/\kappa_B})  +0.23)$ 
         \\
          &Parallel force-mass dipoles& - & $V_{1,x} \sim \frac{-D^2}{\kappa_B\eta h^2}(0.098\log(h\sqrt{T/\kappa_B})  -0.05)$ 
         \\
         &Perpendicular force-mass dipoles& + & $V_{1,x} = 0$  
         \\
         &Force-mass dipoles at $\alpha = \pi/4$& - & $V_{1,x} \sim \frac{D^2}{\kappa_B\eta h^2}(0.333\log(h\sqrt{T/\kappa_B})  +0.49)$ 
         \\
         &Parallel self propulsion& + & $V_{1,x} \sim \text{ slower than} \frac{V_{\rm act}D}{\kappa_B }(\log(h\sqrt{T/\kappa_B}))$ 
         \\
          &Perpendicular self propulsion& + & $V_{1,x}=0$ 
         \\
         &Self propulsion at $\alpha = \pi/4$& - & $V_{1,x} \sim \frac{V_{\rm act}D}{\kappa_B }(0.16\log(h\sqrt{T/\kappa_B})  +0.283)$  
         \\
         \hline
        \end{tabular}
        \caption{Asymptotic results for  velocity correction along $x$ of different singularities}
        \label{tab: excat results x}
    \end{table}

\begin{table}
  [ht] \caption{Result schemes by singularity} \label{tab:stimuli}
  \begin{tabular}
      {|w{c}{6cm}|w{c}{3cm}|w{c}{3cm}|}
      \hline
      Singularity & Deformation scheme & correction type \\
      \hline
      Parallel force dipole $(D_{11})$ & \parbox[c]{8em}{
      \includegraphics[width=1in]{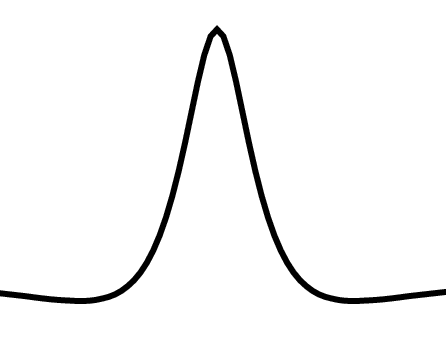}} & enhanced attraction \\
      \hline
      Perpendicular force dipole $(D_{33})$ & \parbox[c]{8em}{
      \includegraphics[width=1in]{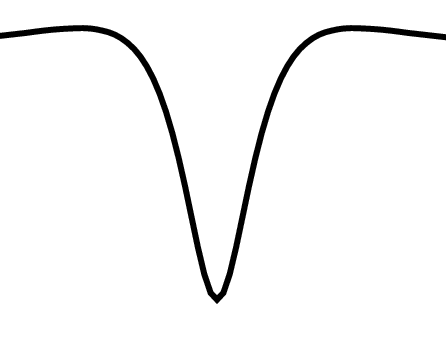}} & reduced repulsion \\
      \hline
      Off diagonal terms $(D_{13},D_{31})$& \parbox[c]{8em}{
      \includegraphics[width=1in]{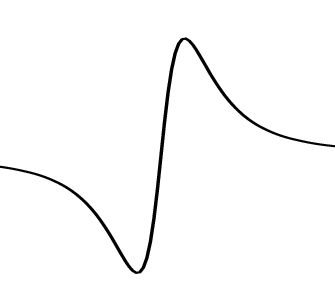}} & generating attraction \\
       \hline
      Parallel mass dipole $(q_1)$& \parbox[c]{8em}{
      \includegraphics[width=1in]{forcexzzx_tau1Scheme.eps}} & generating attraction \\
       \hline
      Perpendicular mass dipole $(q_3)$($\alpha<\pi/2$)& \parbox[c]{8em}{
      \includegraphics[width=1in]{forcepe_tau1_scheme.eps}} & reduced repulsion \\
        \hline
      Perpendicular mass dipole $(q_3)$($\alpha>\pi/2$)& \parbox[c]{8em}{
      \includegraphics[width=1in]{forcepa_tau1_scheme.eps}} & enhanced attraction \\
      \hline
  \end{tabular}
\label{tab: shcematics}
\end{table}

\subsection{Active velocity dominated regime}
\begin{figure}[h]
    \subfloat[]{\includegraphics[width=0.4\linewidth]{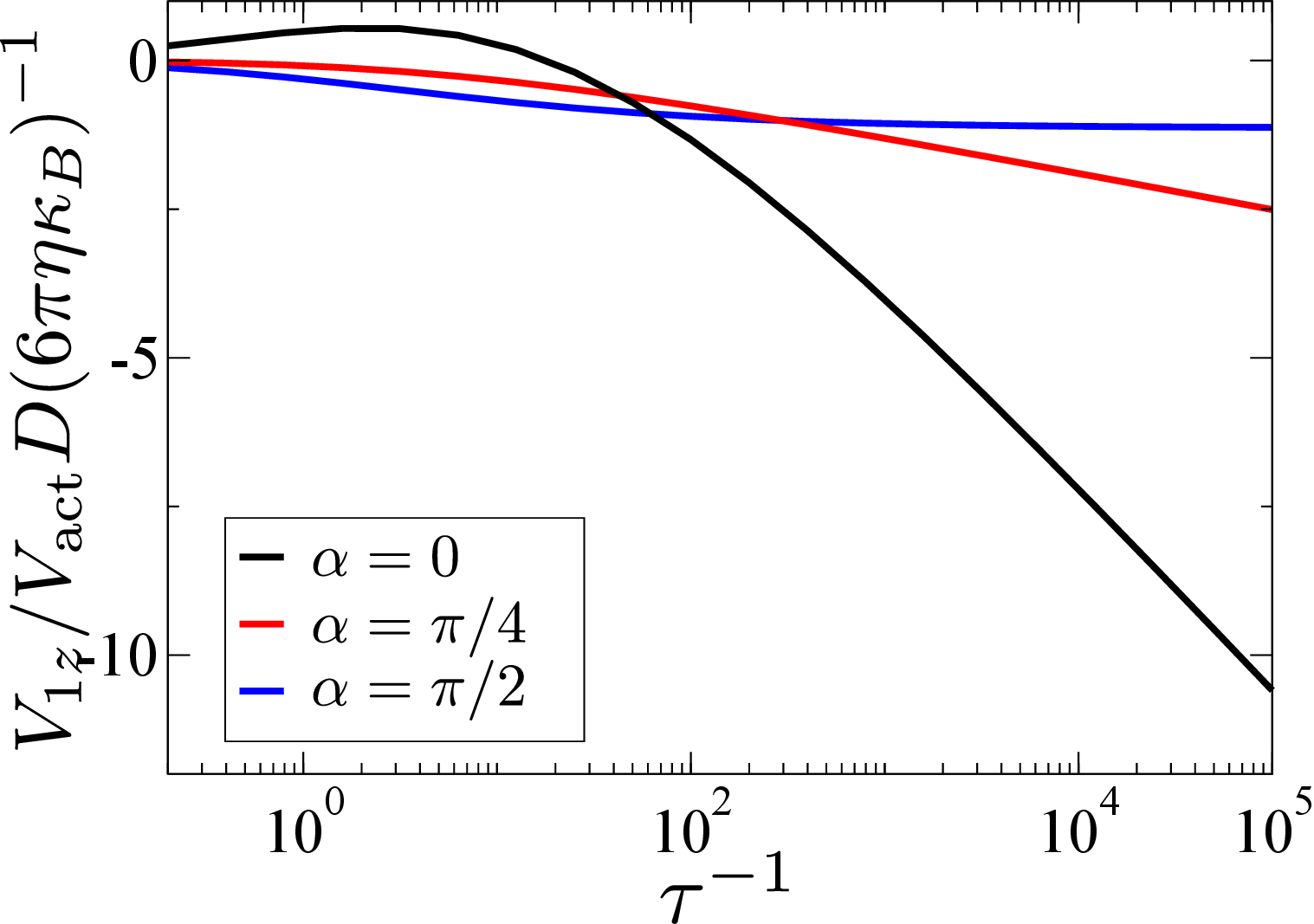}}
    \subfloat[]{\includegraphics[width=0.4\linewidth]{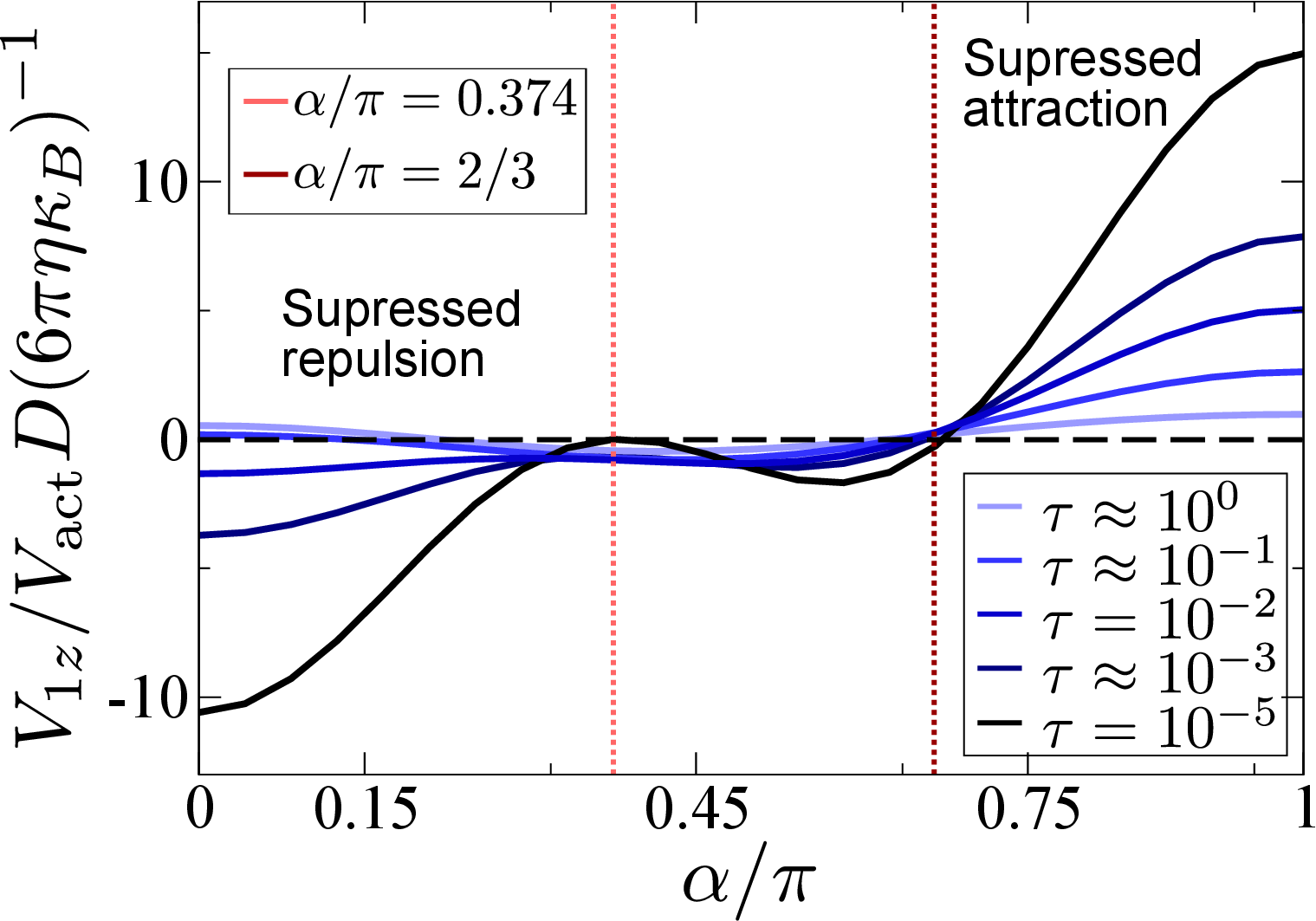}}
    \caption{\small (a) Rescaled induced velocity along $x$ of a self-propelled particle with force and mass dipole combination ($Q =1$) as a function of dimensionless tension $\tau$. Blue line is swimming parallel to the membrane, black line is perpendicular to membrane, and red is swimming in $\alpha = \pi/4$. (b) Rescaled induced velocity along $x$ of a force and mass dipole combination ($Q =1$) as a function of orientation angle $\alpha$}.\label{fig. total V in z}
\end{figure}

\begin{figure}[h]
    \subfloat[]{\includegraphics[width=0.4\linewidth]{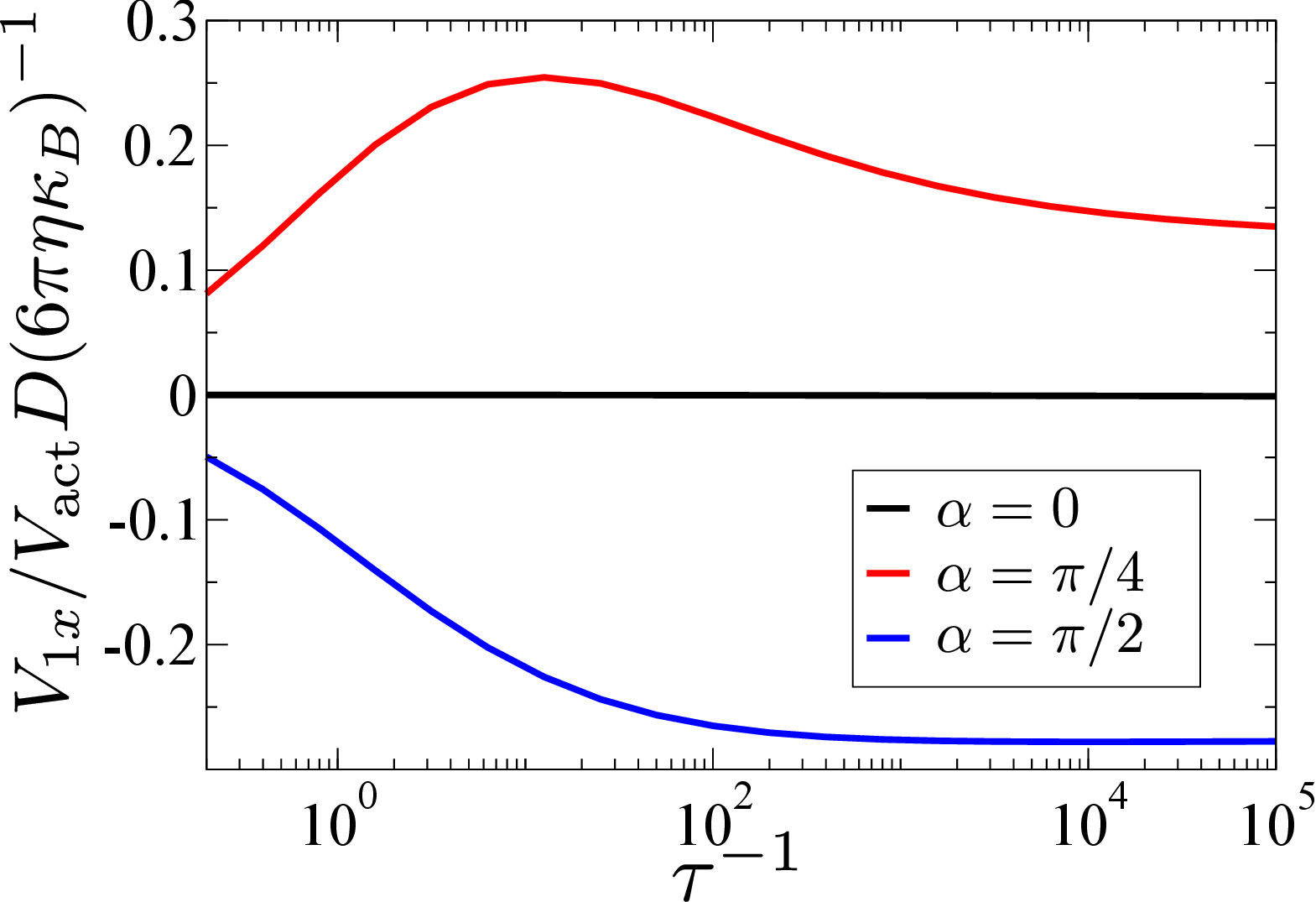}}
    \subfloat[]{\includegraphics[width=0.4\linewidth]{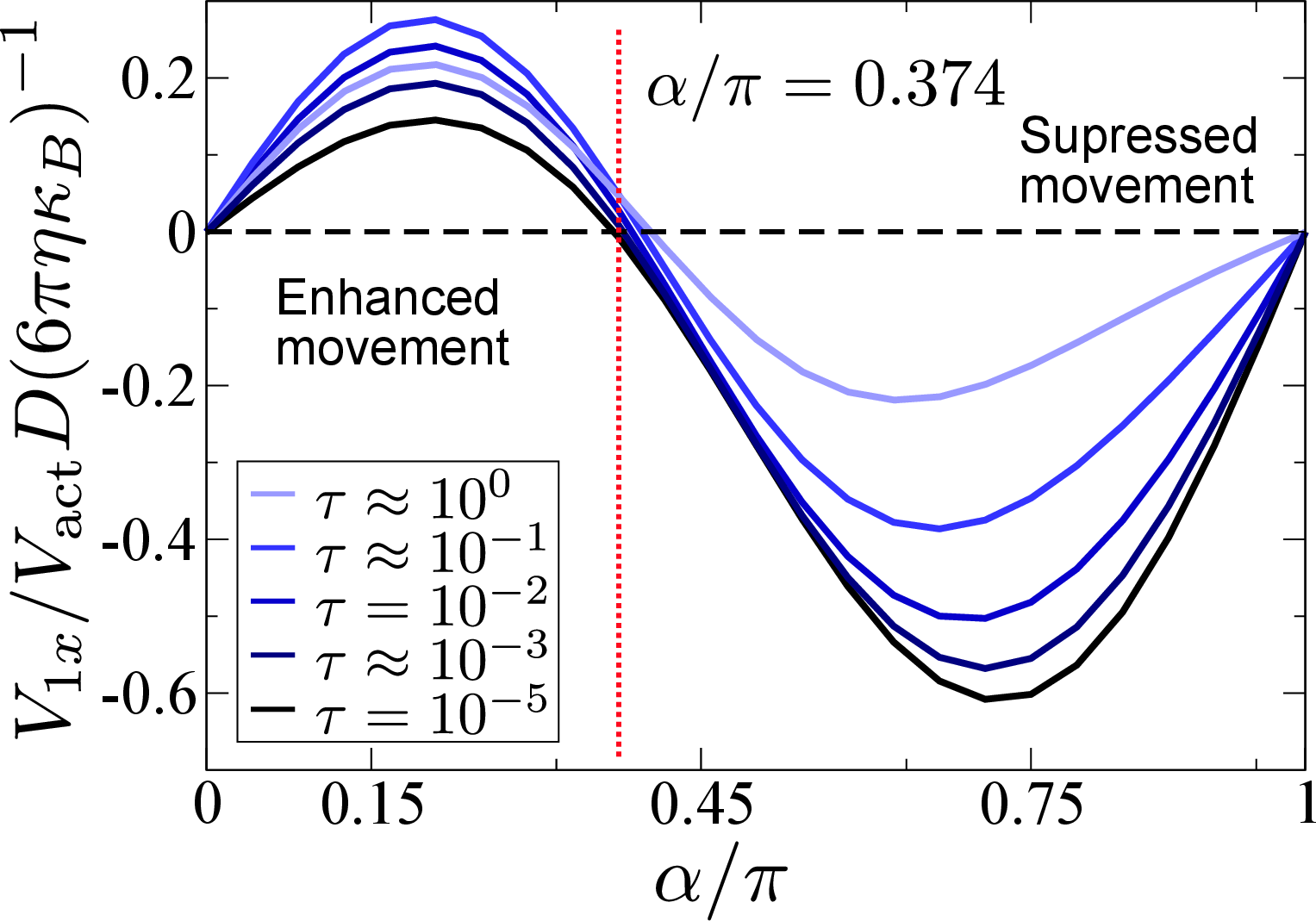}}
    \caption{\small (a) Rescaled induced velocity along $x$ of a self-propelled particle with force and mass dipole combination ($Q =1$) as a function of dimensionless tension $\tau$. Blue line is swimming parallel to the membrane, black line is perpendicular to membrane, and red is swimming in $\alpha = \pi/4$. (b) Rescaled induced velocity along $x$ of a force and mass dipole combination ($Q =1$) as a function of orientation angle $\alpha$}.\label{fig. total V in x}
\end{figure}

Here we present figures for the case where the active velocity dominates significantly over the singularity contributions specifically, $|V_{\rm act}| \gg \frac{|D|}{\eta h_0^2}$.
We set 
\begin{equation}
    \frac{D}{\eta a^2} = 35|V_{\ \rm act}|,  \frac{h_0}{a} = 50\text{ and } Q=1.
\end{equation}
The results are shown in Figs. \ref{fig. total V in z} and \ref{fig. total V in x}.
Notice that the velocity correction along $z$ is not much different than what is presented in Sec \ref{sec self prop}. 
The magnitude of the correction is smaller as expected since the particle is essentially further away from the wall. 
On the other hand, along $x$ there are more pronounced differences compared to Sec \ref{sec self prop}.
Specifically, for small incident angles, the particle's movement is enhanced in the positve $x$ direction.

\subsection{Dominant Force Dipole ($Q \to 0$)}\label{subsection: force}
We now explore the different limits of the dipole ratio $Q$. 
First, consider the limit where $Q \rightarrow0$, where only a stresslet remains. 

\begin{figure}[h]
    \centering
    \subfloat[]{\includegraphics[width=0.4\linewidth]{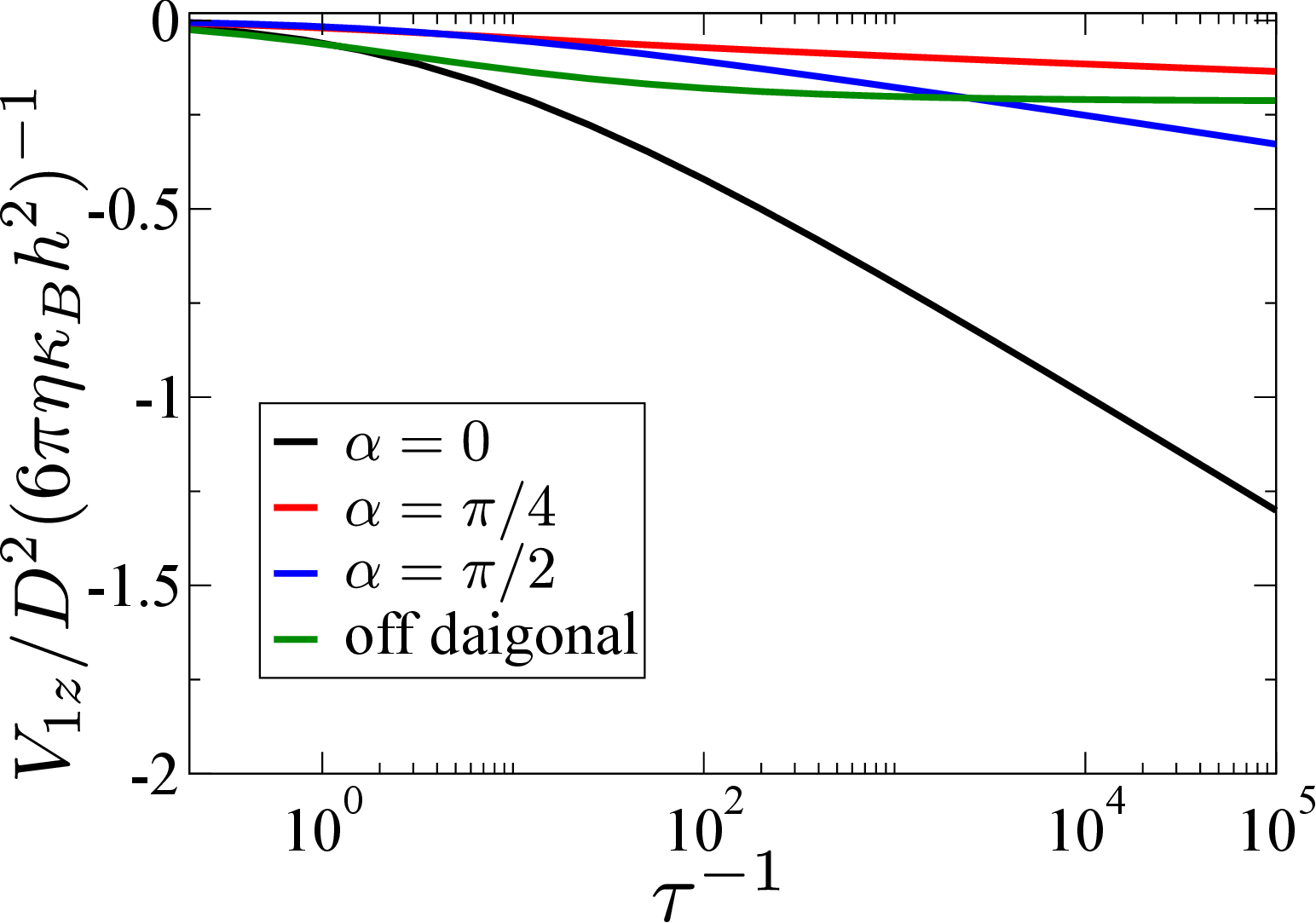}}
    \subfloat[]{\includegraphics[width=0.4\linewidth]{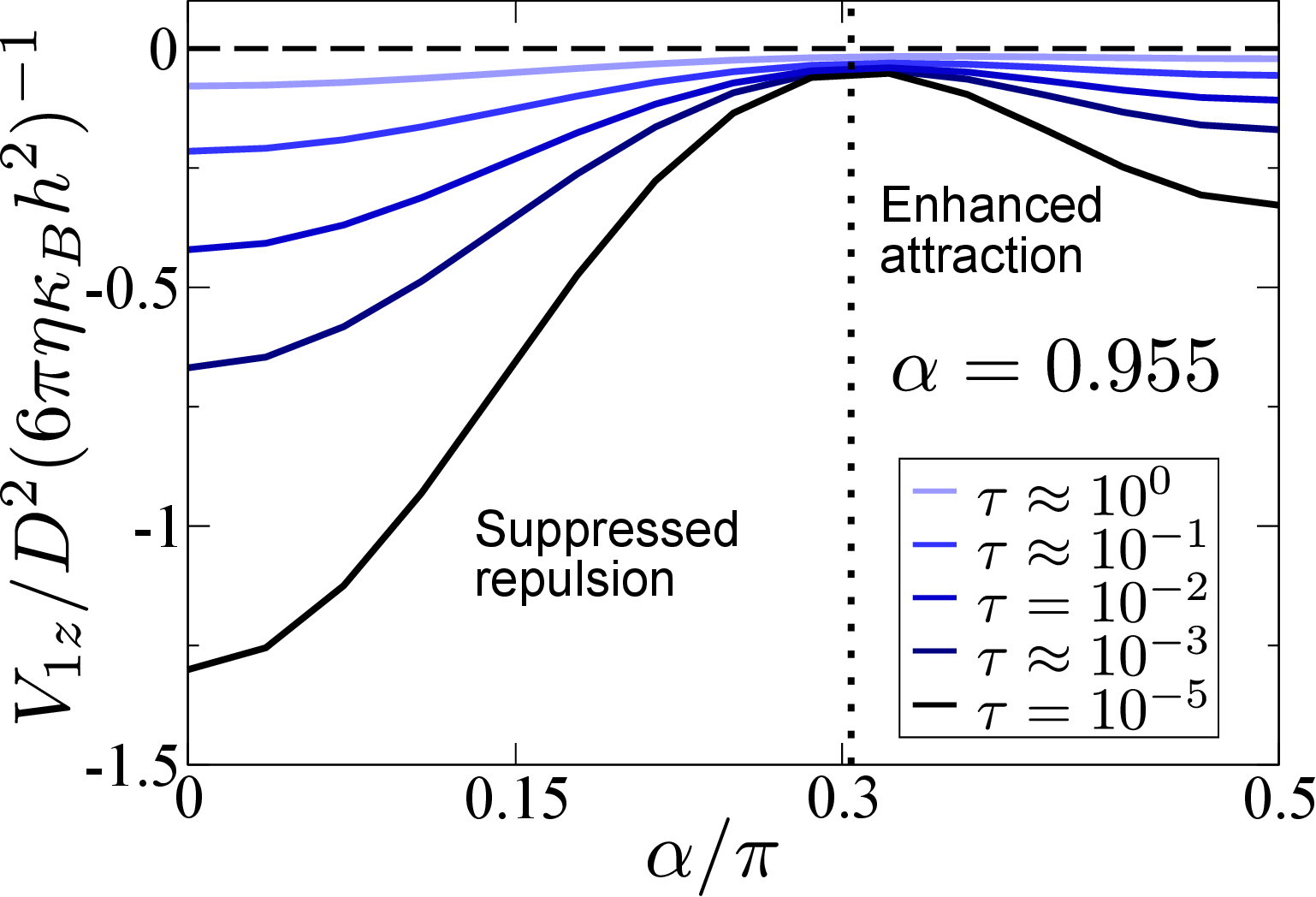}}
    \caption{\small (a) Rescaled induced velocity along $z$ of a stresslet ($Q=0$) as a function of dimensionless tension $\tau$. Blue line is swimming parallel to the membrane, black line is perpendicular to membrane, green line is off diagonal terms only, and red is swimming in $\alpha = \pi/4$ which includes contribution from all components. (b) Rescaled induced velocity along $z$ of a stresslet as a function of orientation angle $\alpha$.\label{fig. forces in z}}
\end{figure}

\textbf{Corrections to the normal velocity.} The velocity correction along $z$ as a function of dimensionless tension $\tau$ is shown in Fig. \ref{fig. forces in z}(a).  
For swimming with parallel, perpendicular and $\alpha =\pi/4$ orientations, the velocity scales as $\sim\log(\tau^{-1})$ while for the off-diagonal terms the velocity correction is even smaller (see table \ref{tab: excat results z} for more details). 
The dependence on $\alpha$ can be understood qualitatively using similar arguments as before. A stresslet is repelled from a rigid wall for small angles, and is attracted to it for large angles. The crossover between these regimes is given by setting $Q = 0$ in Eq. (\ref{eq cross total}) and found to be about $\alpha_{\rm cross} = 0.304\pi \approx 0.955$ (dotted vertical line in fig \ref{fig. forces in z}(b)). For $\alpha < \alpha_{\rm cross}$, the repulsion is suppressed since the membrane deforms away from the particle, while for $\alpha>\alpha_{\rm cross}$, the attraction is enhanced since the membrane is pulled towards the particle. From symmetry, the dynamics along $z$ are identical for $\alpha \in (\pi/2, \pi)$ as for $\alpha \in [0, \pi/2]$.


\begin{figure}[h]
    \centering
    \subfloat[]{\includegraphics[width=0.4\linewidth]{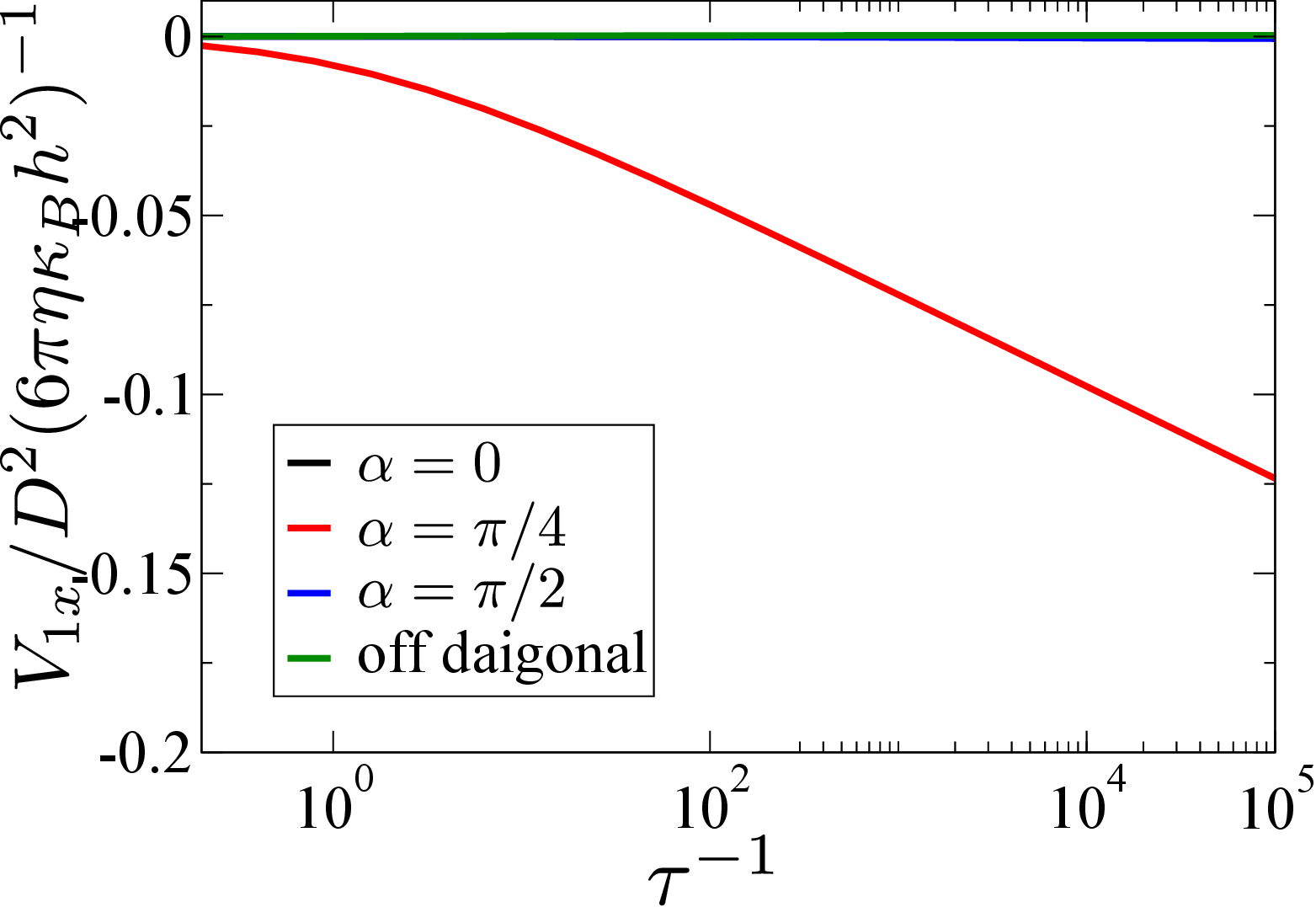}}
    \subfloat[]{\includegraphics[width=0.4\linewidth]{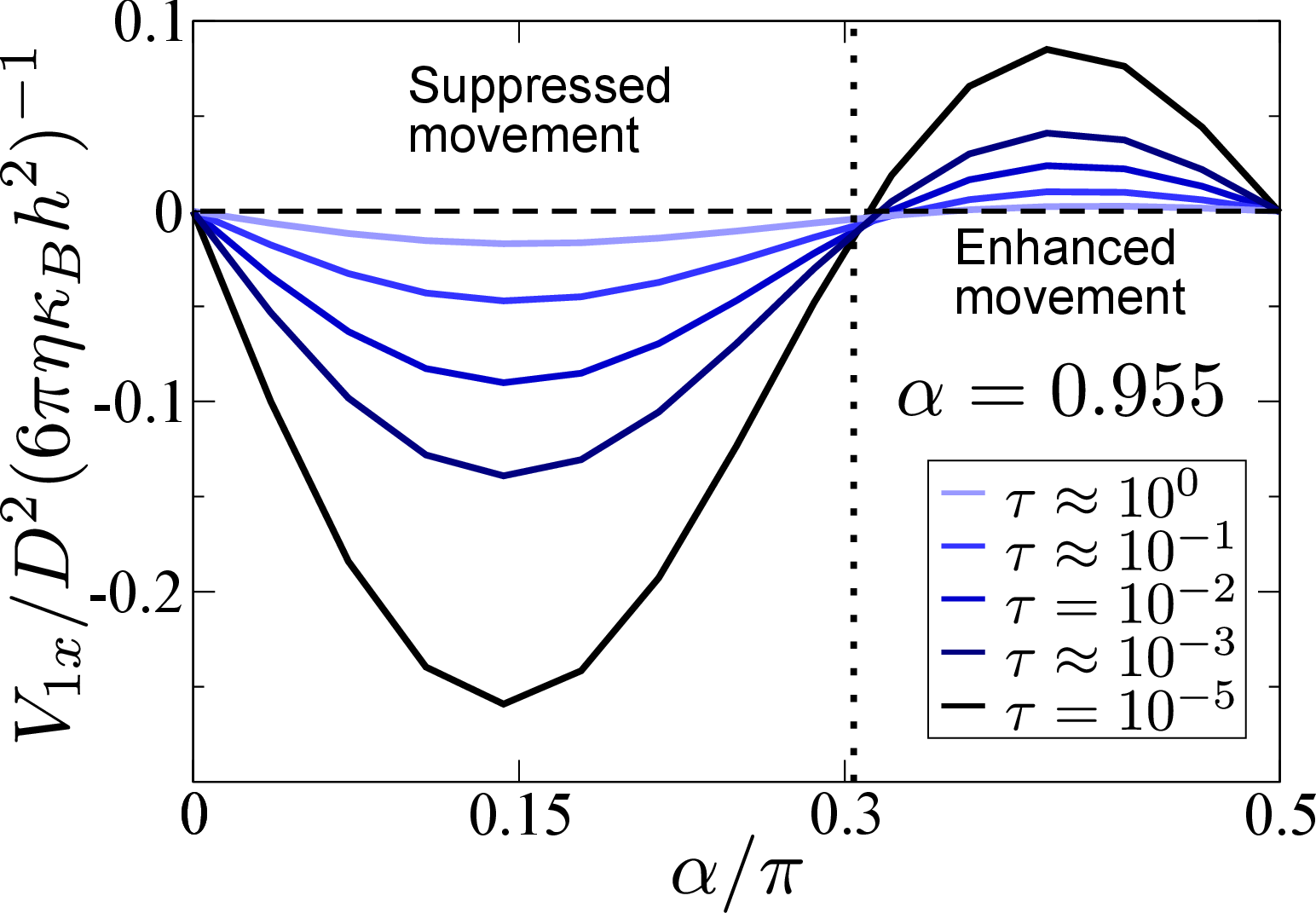}}
    \caption{\small (a) Rescaled induced velocity along $x$ of a stresslet ($Q = 0$) as a function of dimensionless tension $\tau$. Blue line is swimming parallel to the membrane, black line is perpendicular, green line is off diagonal terms only, and red is swimming at $\alpha = \pi/4$. (b) Rescaled induced velocity along $x$ of a stresslet ($Q = 0$) as a function of orientation angle $\alpha$.\label{fig. forces in x of theta}}
\end{figure}

\textbf{Corrections to the tangential velocity.} 
Fig. \ref{fig. forces in x of theta}(a) shows the velocity correction along $x$ of the different components of a stresslet as a function of the dimensionless tension $\tau$. 
Notice that due to the symmetry of the problem, for each component separately the velocity correction is zero while the correction at $\alpha = \pi/4$ scales as $\sim\log(\tau^{-1})/h^2$. 
Fig. \ref{fig. forces in x of theta}(b) shows that for $\alpha \in(0,\alpha_{\rm cross})$ the deformation suppresses motion in the positive $x$ direction, while for $\alpha \in(\alpha_{\rm cross},\pi/2)$ the deformation enhances the motion.
Lastly, we note that from a symmetry argument for $\alpha \in (\pi/2,\pi]$, we have that $V_{1,x} (\alpha) = -V_{1,x} (\pi -\alpha)$. Since in that range of $\alpha$ the velocity $V_0$ is along negative $x$, the dynamics change sign. 

\subsection{Dominant Mass Dipole ($Q \rightarrow\infty)$}\label{subsection: source}
We consider now the limit $Q \rightarrow\infty$ where only the mass dipole remains. 
Since now $D = 0$, a new velocity scale for the problem is chosen, specifically, $V_P = q(\eta h_0^3)^{-1}$ leading to $\Lambda = q(h_0\kappa_B)^{-1}$.
Therefore, the correction velocity due to deformation now scales as $V_1 \sim \frac{q^2}{\eta \kappa_B h^4}$. In this case, the crossover angle according to \eqref{eq cross total} is $\alpha_{cross} = \pi/2$. 

\begin{figure}[h]
    \centering
    \subfloat[]{\includegraphics[width=0.4\linewidth]{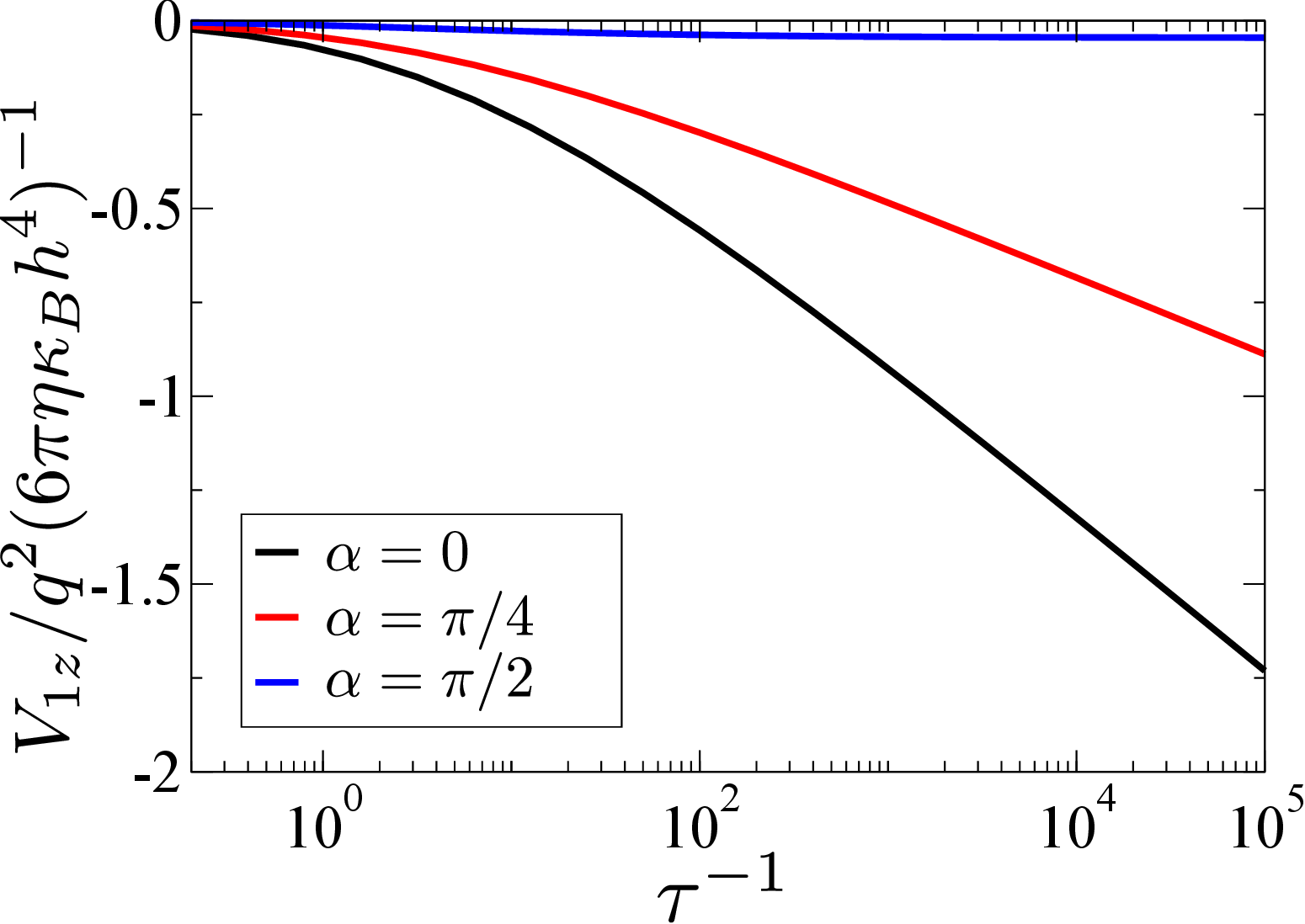}}
    \subfloat[]{\includegraphics[width=0.4\linewidth]{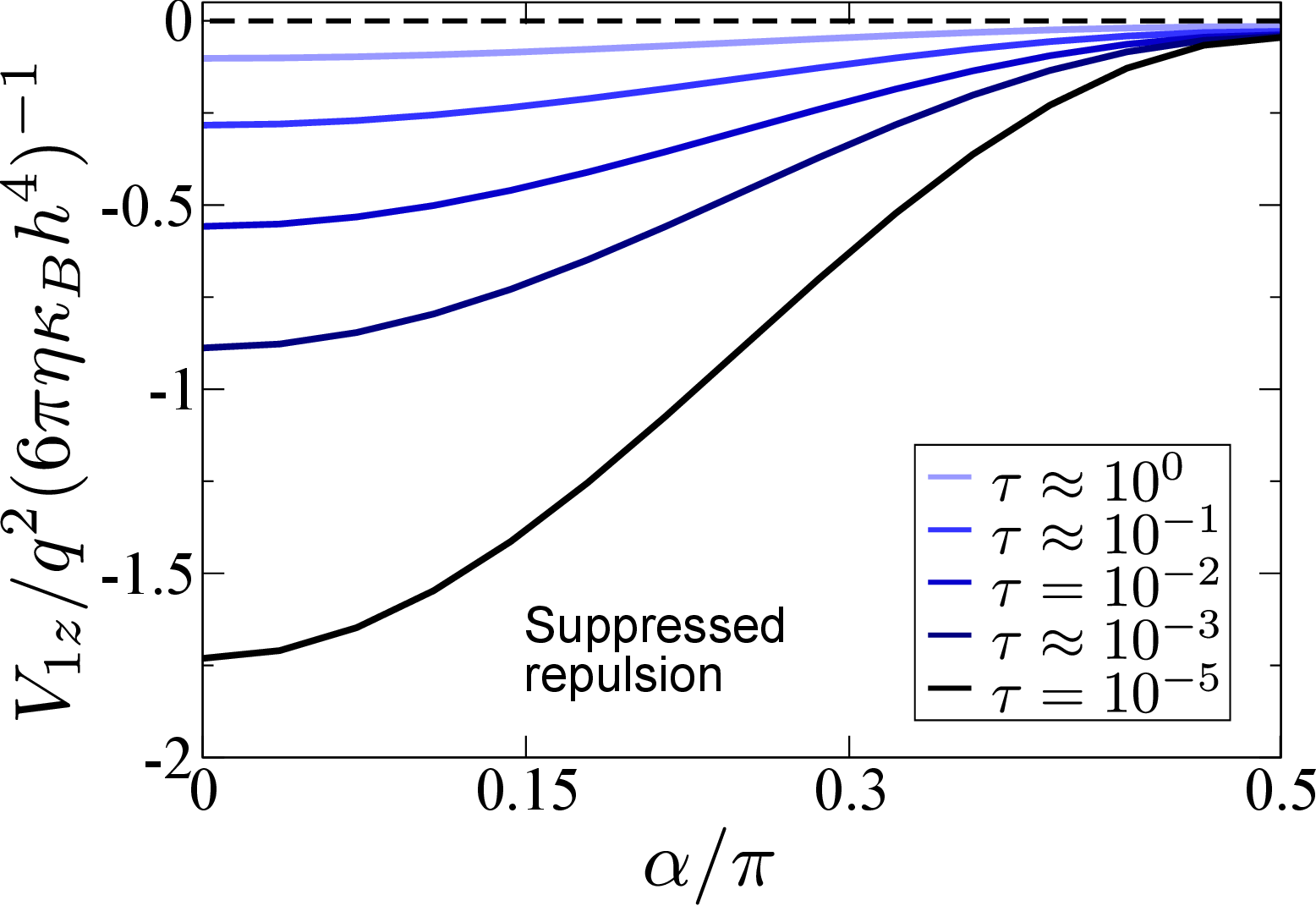}}
    \caption{\small (a) Rescaled induced velocity along $z$ of a mass dipole ($Q\rightarrow\infty$) as a function of dimensionless tension $\tau$. Blue line is swimming parallel to the membrane, black line is perpendicular, and red is swimming at $\alpha = \pi/4$. (b) Rescaled induced velocity along $z$ of a mass dipole as a function of orientation angle $\alpha$.\label{fig. sources in z}}
\end{figure}

\textbf{Corrections to the perpendicular velocity.} The results of the velocity correction along $z$ as a function of dimensionless tension $\tau$ are shown in Fig. \ref{fig. sources in z}(a) where $V_1 \sim \log(\tau^{-1})/h^4$ or slower. 
Notice that the correction due to the parallel term is not identically zero but is much smaller than the correction due to the perpendicular component. 
The correction along $z$ as a function of $\alpha$ is presented in Fig. \ref{fig. sources in z}(b). From Eq. (\ref{eq total velocity}), we see that the particle moves away from the membrane for $0<\alpha < \pi/2$, Deformation suppresses this repulsion.

\begin{figure}[h]
    \centering
    \subfloat[]{\includegraphics[width=0.4\linewidth]{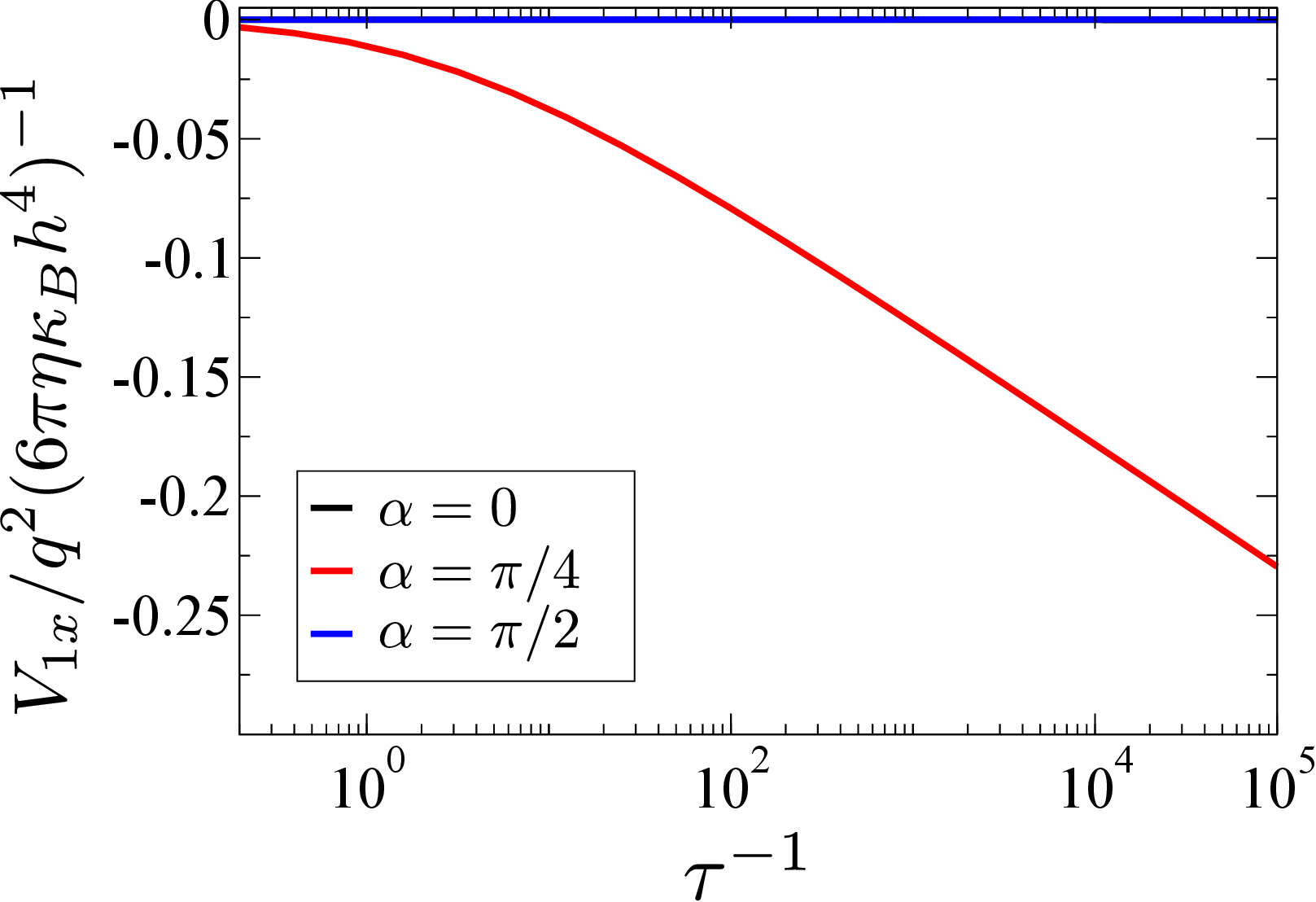}}
    \subfloat[]{\includegraphics[width=0.4\linewidth]{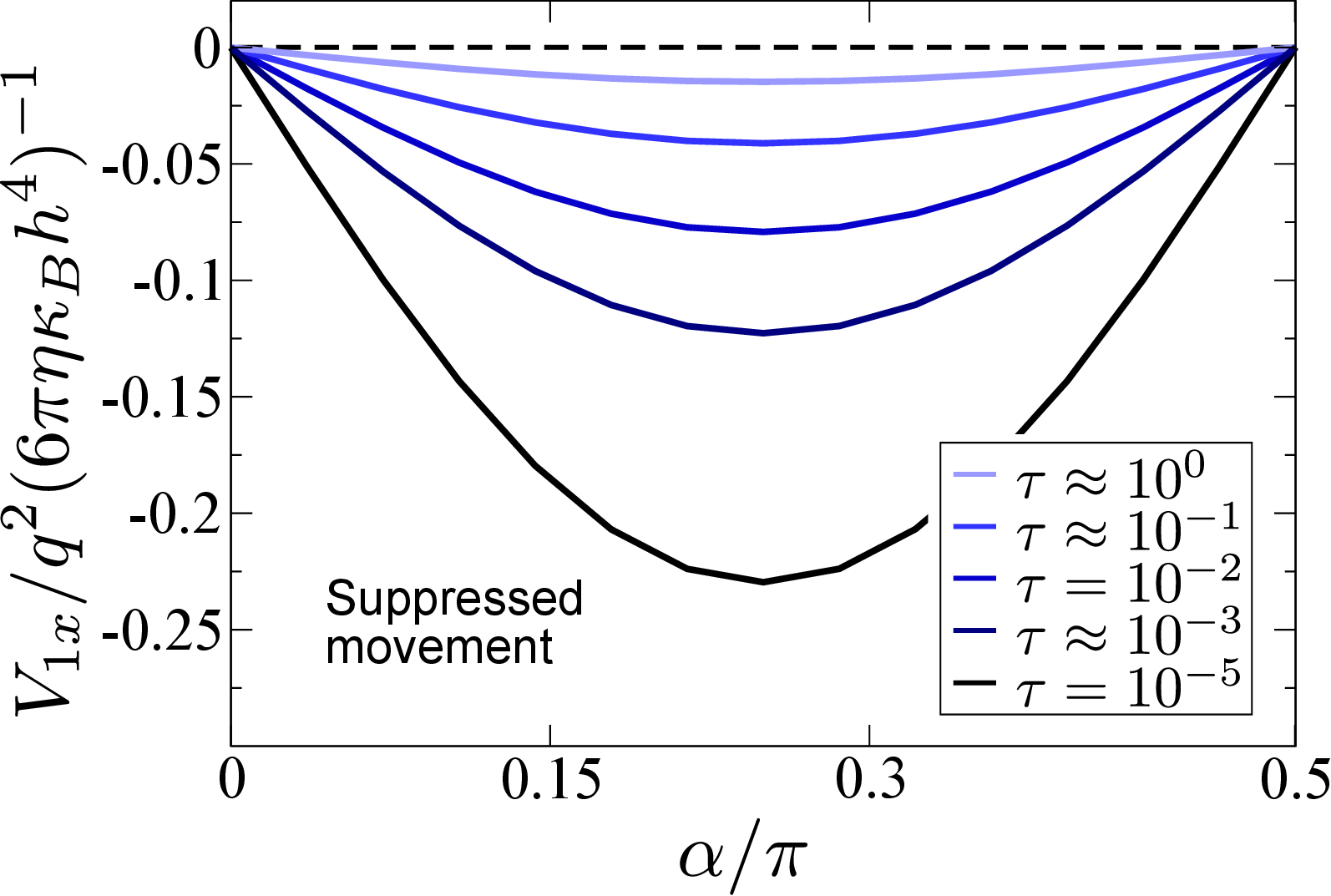}}
    \caption{\small Rescaled induced velocity along $x$ of a mass dipole ($Q\rightarrow\infty$) as a function of dimensionless tension $\tau$. Blue line is swimming parallel to the membrane, black line is perpendicular to membrane, and red is swimming in $\alpha = \pi/4$. (b) Rescaled induced velocity along $x$ of a mass dipole as a function of orientation angle $\alpha$.\label{fig. sources in x}}
\end{figure}

\textbf{Corrections to the parallel velocity.} The results of the correction along $x$ are presented in Figs. \ref{fig. sources in x}(a) and \ref{fig. sources in x}(b); see tables \ref{tab: excat results z} and \ref{tab: excat results x} for more details. For $0<\alpha<\pi$, the movement of the mass dipole near a flat wall is along the positive $x$ direction. Deformation suppresses this movement for $0<\alpha<\pi/2$ and enhances it for $\pi/2<\alpha<\pi$. Once more, this is consistent with \eqref{eq cross total}, which predicts $\alpha_{\rm cross} = \pi/2.$


\end{document}